\documentclass[aps,prd,preprint,superscriptaddress]{revtex4}
\usepackage{graphicx}

\begin{document}
\newcommand{\vecvar}[1]{\mbox{\boldmath$#1$}}


\title{Generalized form factors, generalized parton distributions\\
and the spin contents of the nucleon}


\author{M.~Wakamatsu and Y.~Nakakoji}
\affiliation{Department of Physics, Faculty of Science, \\
Osaka University, \\
Toyonaka, Osaka 560-0043, JAPAN}



\begin{abstract}
With a special intention of clarifying the underlying spin contents
of the nucleon, we investigate the generalized form factors
of the nucleon, which are defined as the $n$-th $x$-moments of the
generalized parton distribution functions, within the
framework of the chiral quark soliton model.
A particular emphasis is put on the pion mass dependence
of final predictions, which we shall compare with the
predictions of lattice QCD simulations carried out in the
so-called heavy pion region around $m_\pi \simeq (700 \sim 900)
\,\mbox{MeV}$. We find that some observables are very sensitive
to the variation of the pion mass.
It will be argued that the negligible importance
of the quark orbital angular momentum indicated by the LHPC
and QCDSF lattice collaborations might be true in the unrealistic
heavy pion world, but it is not necessarily the case in our real
world close to the chiral limit.
\end{abstract}


\maketitle


\section{Introduction}
The so-called ``nucleon spin crisis'' raised by
the European Muon Collaboration (EMC) measurement in 1988 is
one of the most outstanding findings in the field of hadron
physics \cite{EMC88},\cite{EMC89}.
The renaissance of the physics of high energy deep inelastic
scatterings is greatly indebted to this epoch-making finding.
Probably, one of the most outstanding progresses achieved recently in
this field of physics is the discovery and the subsequent research
of completely new observables called generalized parton distribution
functions (GPD). It has been revealed that the GPDs, which can be
measured through the so-called deeply-virtual Compton scatterings
(DVCS) or the deeply-virtual meson productions (DVMP), contain
surprisingly richer information than the standard parton distribution
functions \cite{MRGDH94}--\cite{RP02}. 

Roughly speaking, the GPDs are generalization of ordinary parton
distributions and the elastic form factors of
the nucleon. The GPDs in the most general form are functions of three
kinematical variables : the average longitudinal momentum fraction
$x$ of the struck parton in the initial and final states,
a skewdness parameter $\xi$ which measures the difference between
two momentum fractions, and the four-momentum-transfer square $t$ of
the initial and final nucleon. In the forward limit $t \rightarrow 0$,
some of the GPDs reduce to the usual quark, antiquark and gluon
distributions. On the other hand, taking the $n$-th moment of
the GPDs with respect to the variable $x$, one obtains the
generalizations of the electromagnetic form factors of the nucleon,
which are called the generalized form factors of the nucleon.
The complex nature of the GPDs, i.e. the fact that they are functions
of three variable, makes it quite difficult to grasp their
full characteristics both experimentally and theoretically.
From the theoretical viewpoint, it may be practical to begin studies
with the two limiting cases. The one is the forward limit of
zero momentum transfer. We have mentioned that, in this limit, some of
the GPDs reduce to the ordinary parton distribution function
depending on one variable $x$. However, it turns out that, even
in this limit, there appear some completely new distribution
functions, which cannot be accessed by the ordinary inclusive
deep-inelastic scattering measurements.
Very interestingly, it was shown by Ji that one of such
distributions contains valuable information on the total angular
momentum carried by the quark fields in the
nucleon \cite{Ji97}--\cite{JTH96}. This information,
combined with the known knowledge on the longitudinal quark
polarization, makes it possible to determine the quark orbital
angular momentum contribution to the total nucleon spin purely
experimentally.

Another relatively-easy-to-handle quantities are the
generalized form factors of the nucleon \cite{DFJK05},\cite{Diehl05},
which are given as the
non-forward nucleon matrix elements of the spin-$n$, twist-two
quark and gluon operators. Since these latter quantities are given
as the nucleon matrix elements of local operators, they can be
objects of lattice QCD simulations.
(It should be compared with parton distributions. The direct
calculation of parton distributions is beyond the scope of lattice
QCD simulations, since it needs to treat the nucleon matrix elements
of quark bilinears, which are {\it nonlocal in time}.)
In fact, two groups, the LHPC collaboration and the QCDSF collaboration
independently investigated the generalized form factors of the nucleon,
and gave several interesting predictions, which can in principle be
tested by the measurement of GPDs in the
near future \cite{LHPC02}--\cite{QCDSF05}.
Although interesting, there is no {\it a priori} reason to believe
that the predictions of these lattice simulations are realistic
enough. The reason is mainly that the above mentioned lattice
simulation were carried out in the heavy pion regime around
$m_\pi \simeq (700 \sim 900) \,\mbox{MeV}$  with neglect
of the so-called disconnected diagrams. Our real world is rather
close to the chiral limit with vanishing pion mass, and we know that,
in this limit, the Goldstone pion plays very important roles in
some intrinsic properties of the nucleon. The lattice simulation
carried out in the heavy pion region is in danger of missing some
important role of chiral dynamics. 

On the other hand, the chiral quark soliton model (CQSM) is an
effective model of baryons, which maximally incorporates the chiral
symmetry of QCD and its spontaneous breakdown \cite{DPP88},\cite{WY91}.
(See \cite{Wakamatsu92R}--\cite{ARW96} for early reviews.) It was already
applied to the physics of ordinary parton distribution functions
with remarkable success \cite{WGR96}--\cite{Wakamatsu03b}.
For instance, an indispensable role of
pion-like quark-antiquark correlation was shown to be essential to
understand the famous NMC measurement, which revealed the dominance
of the $\bar{d}$-quark over the $\bar{u}$-quark inside the
proton \cite{WK98},\cite{PPGWW99},\cite{Wakamatsu92}.
Then, it would be interesting to see what predictions the CQSM would
give for the quantities mentioned above. Now, the main purpose of the
present study is to study the generalized form factors of the
nucleon within the framework of the CQSM and compare
its predictions with those of the lattice QCD simulations.
Of our particular interest  here is to see the change of
final theoretical predictions against the variation of the pion mass.
Such an analysis is expected to give some hints for judging the
reliability of the lattice QCD predictions at the present level
for the observables in question.

The plan of the paper is as follows. In sect.II, we shall briefly
explain how to introduce the nonzero pion mass into the scheme
of the CQSM with Pauli-Villars regularization.
In sect.III, we derive the theoretical expressions for the
generalized form factors of the nucleon.
Sect.IV is devoted to the discussion of the results of the numerical
calculations. Some concluding remarks are then given in sect.V.

\section{Model lagrangian with pion mass term}

We start with the basic effective lagrangian of the chiral quark soliton 
model in the chiral limit given as
\begin{equation}
 {\cal L}_0 = \bar{\psi} (x) (i \not\!\partial - M U^{\gamma_5} (x) ) 
 \psi (x) ,
\end{equation}
with 
\begin{equation}
 U^{\gamma_5} (x) \ = \ 
 e^{i \gamma_5 \vecvar{\tau} \cdot \vecvar{\pi} (x) / f_{\pi}}
 \ = \ \frac{1 + \gamma_5}{2} \,U (x) + 
 \frac{1 - \gamma_5}{2} \,U^{\dagger} (x) .
\end{equation}
which describes the effective quark fields, with a dynamically generated 
mass $M$, strongly interacting with pions \cite{DPP88},\cite{WY91}.
Since one of the main purposes of the 
present study is to see how the relevant observables depend on pion mass, 
we add to ${\cal L}_0$ an explicit chiral symmetry breaking term 
${\cal L}^{\prime}$ given by \cite{KWW99}
\begin{equation}
 {\cal L}^{\prime} = \frac{1}{4} \,f_{\pi}^2 \,m_{\pi}^2 \,\mbox{tr}_f
 \,[ U (x) + U^{\dagger} (x) - 2] . \label{eq:LSB}
\end{equation}
Here the trace in (\ref{eq:LSB}) is to be taken with respect to
flavor indices. 
The total model lagrangian is therefore given by
\begin{equation}
 {\cal L}_{CQM} = {\cal L}_0 + {\cal L}^{\prime} .
\end{equation}
Naturally, one could have taken an alternative choice in which the
explicit chiral-symmetry-breaking effect is introduced in the form of
current quark mass term as ${\cal L}^{\prime} = -m_0 \bar{\psi} \psi$.
We did not do so, because we do not know any consistent regularization
of such effective lagrangian with finite current quark mass within the 
framework of the Pauli-Villars subtraction scheme, as explained in
Appendix of \cite{KWW99}. 
The effective action corresponding to the above lagrangian is given as
\begin{equation}
 S_{eff} [U] = S_F [U] + S_M [U] , \label{eq:EnergSol}
\end{equation}
with
\begin{equation}
 S_F [U] = - \,i \,N_c \,\mbox{Sp} \,
 \ln ( i \not\!\partial - M U^{\gamma_5} ) ,
\end{equation}
and
\begin{equation}
 S_M [U]  = \int \frac{1}{4} \,f_{\pi}^2 \,m_{\pi}^2 \,
 \mbox{tr}_f \,[ U (x) + U^{\dagger} (x) - 2 ] \,d^4 x .
\end{equation}
Here $\mbox{Sp} \hat{O} = \int d^4 x \,\mbox{tr}_\gamma \,\mbox{tr}_f \,
\langle x | \hat{O} | x \rangle$ with $\mbox{tr}_\gamma$ and
$\mbox{tr}_f$ representing the trace of the Dirac gamma matrices and
the flavors (isospins), respectively.
The fermion (quark) part of the above action contains ultra-violet 
divergences. To remove these divergences, we must introduce physical
cutoffs. For the purpose of regularization, here we use the
Pauli-Villars subtraction scheme. As explained in \cite{KWW99},
we must eliminate not
only the logarithmic divergence contained in $S_f [U]$ but also
the quadratic and 
logarithmic divergence contained in the equation of motion shown below.
To get rid of all these troublesome divergence, we need at least two 
subtraction terms. The regularized action is thus defined as 
\begin{equation}
 S_{eff}^{reg} [U] = S_F^{reg} [U] + S_M [U] ,
\end{equation}
where
\begin{equation}
 S_F^{reg} [U] = S_F [U] - \sum_{i = 1}^2 \,c_i \,S_F^{\Lambda _i} [U] .
\end{equation}
Here $S_F^{\Lambda_i}$ is obtained from $S_F [U]$ with $M$ replaced by
the Pauli-Villars regulator mass $\Lambda_i$. These parameters are fixed
as follows.
First, the quadratic and logarithmic divergence contained in the
equation of motion (or in the expression of the vacuum quark condensate) 
can, respectively, removed if the subtraction constants satisfy the 
following two conditions :
\begin{eqnarray}
 M^2 &-& \sum_{i = 1}^2 c_i \Lambda_i^2 = 0 ,\\
 M^4 &-& \sum_{i = 1}^2 c_i \Lambda_i^4 = 0 .
\end{eqnarray}
(We recall that the condition which removes the logarithmic divergence 
in $S_f [U]$ just coincides with the 1st of the above conditions.)
By solving the above equations for $c_1$ and $c_2$, we obtain
\begin{eqnarray}
 c_1 &=& \left(\frac{M}{\Lambda_1} \right)^2 
 \frac{\Lambda_2^2 - M^2}{\Lambda_2^2 - \Lambda_1^2} , \\
 c_2 &=& - \left(\frac{M}{\Lambda_2} \right)^2 
 \frac{\Lambda_1^2 - M^2}{\Lambda_2^2 - \Lambda_1^2} ,
\end{eqnarray}
which constrains the values of $c_1$ and $c_2$, once $\Lambda_1$ and 
$\Lambda_2$ are given. For determining $\Lambda_1$ and $\Lambda_2$, 
we use two conditions
\begin{equation}
 \frac{N_c M^2}{4 \pi^2} \,\sum_{i = 1}^2 \,c_i \,
 \left(\frac{\Lambda_i}{M} \right)^2 \,
 \ln \left(\frac{\Lambda_i}{M} \right)^2 = f_{\pi}^2 ,
\end{equation}
and 
\begin{equation}
 \langle \bar{\psi} \psi \rangle _{vac} = \frac{N_c M^3}{2 \pi^2} \,
 \sum_{i = 1}^2 \,c_i \,\left(\frac{\Lambda_i}{M} \right)^4 
 \ln \left(\frac{\Lambda_i}{M} \right)^4 ,
\end{equation}
which amounts to reproducing the correct normalization of the pion
kinetic term in the effective meson lagrangian and also the empirical
value of the vacuum quark condensate.
To derive soliton equation of motion, we must first write down a 
regularized expression for the static soliton energy.
Under the hedgehog ansatz
$\vecvar{\pi} (\vecvar{x}) = f_{\pi} \hat{\vecvar{r}} F (r)$
for the background pion fields, it is obtained in the form : 
\begin{equation}
 E_{static}^{reg} [F (r)] = E_F^{reg} [F (r)] + E_M [F (r)] ,
\end{equation}
where the meson (pion) part is given by
\begin{equation}
 E_M [F (r)] = - \,f_{\pi}^2 \,m_{\pi}^2 \int d^3 x \,[\cos F (r) - 1] ,
\end{equation}
while the fermion (quark) part is given as 
\begin{equation}
 E_F^{reg} [F (r)] = E_{val} + E_{v p}^{reg} , \label{eq:Estatic}
\end{equation}
with
\begin{eqnarray}
 E_{val} &=& N_c \,E_0 , \\
 E_{v p}^{reg} &=& N_c \,\sum_{n < 0} \,(E_n - E_n^{(0)})
 - \sum_{i = 1}^N \,c_i \,N_c \,\sum_{n < 0} \,(E_n^{\Lambda_i} 
 - E_n^{(0) \Lambda_i}) .\label{eq:regenerg}
\end{eqnarray}
Here $E_n$ are the quark single-particle energies, given as the 
eigenvalues of the static Dirac Hamiltonian in the background pion
fields : 
\begin{equation}
 H \,| n \rangle = E_n | n \rangle ,
\end{equation}
with 
\begin{equation}
 H = \frac{\vecvar{\alpha} \cdot \nabla}{i} + \beta M \,
 [ \cos F(r) + i \gamma_5 \vecvar{\tau} \cdot 
 \hat{\vecvar{r}} \sin F(r)] , \label{eq:DiracH}
\end{equation}
while $E_n^{(0)}$ denote energy eigenvalues of the vacuum Hamiltonian
given by (\ref{eq:DiracH}) with $F (r) = 0$ (or $U = 1$).
The particular state $| n = 0 \rangle$, which is a discrete
bound-state orbital coming from the upper Dirac continuum under the
influence of the hedgehog mean field, is called the valence level.
The symbol $\sum_{n < 0}$ in (\ref{eq:regenerg}) denotes the
summation over all the negative energy eigenstates of $H$, i.e.
the negative energy Dirac continuum.
The soliton equation of motion is 
obtained from the stationary condition of $E_{static}^{reg} [F (r)]$
with respect to the variation of the profile function $F (r)$ : 
\begin{eqnarray}
 0 &=& \frac{\delta}{\delta F (r)} \,E_{static} [F (r)] \nonumber \\
 &=& 4 \pi r^2 \,\left\{ - \,M \,[S (r) \sin F (r) - P (r) \cos F (r)]
 + f_{\pi}^2 m_{\pi}^2 \sin F (r) \right\} ,
\end{eqnarray}
which gives 
\begin{equation}
 F (r) = \arctan 
 \left(\frac{P (r)}{S (r) - f_{\pi}^2 m_{\pi}^2 / M} \right) .
 \label{eq:profile}
\end{equation}
Here $S (r)$ and $P (r)$ are regularized scalar and pseudoscalar quark 
densities given as 
\begin{eqnarray}
 S (r) &=& S_{val} (r) + \sum_{n < 0} S_n (r) 
 - \sum_{i = 1}^2 \,c_i \,\frac{\Lambda_i}{M} \,\sum_{n < 0} \,
 S_n^{\Lambda_i} (r) , \\
 P (r) &=& P_{val} (r) + \sum_{n < 0} P_n (r) 
 - \sum_{i = 1}^2 \,c_i \,\frac{\Lambda_i}{M} \,\sum_{n < 0} \,
 P_n^{\Lambda_i} (r) ,
\end{eqnarray}
with
\begin{eqnarray}
 S_n (r) &=& \frac{N_c}{4 \pi} \int d^3 x \,\,
 \langle n | \vecvar{x} \rangle \,
 \gamma^0 \,\frac{\delta (| \vecvar{x}| - r )}{r^2} \,
 \langle \vecvar{x} | n \rangle , \\
 P_n (r) &=& \frac{N_c}{4 \pi} \int d^3 x \,
 \langle n | \vecvar{x} \rangle \,i \,
 \gamma^0 \,\gamma_5 \,\vecvar{\tau} \cdot \hat{\vecvar{r}} \,
 \frac{\delta (| \vecvar{x}| - r )}{r^2}
 \langle \vecvar{x} | n \rangle ,
\end{eqnarray}
while $S_n^{\Lambda_i} (r)$ and $P_n^{\Lambda_i} (r)$ are the
corresponding densities evaluated with the regulator mass
$\Lambda_i$ instead of the dynamical quark mass $M$.
We also note that $S_{val} (r) \equiv S_{n=0} (r)$ and
$P_{val} (r) \equiv P_{n=0} (r)$.
As usual, a self-consistent soliton solution is obtained as follows
with use of Kahana and Ripka's discretized momentum
basis \cite{KR84},\cite{KRS84}.
First by assuming an appropriate (though arbitrary) soliton profile
$F (r)$, the eigenvalue problem of the Dirac Hamiltonian is solved.
Using the resultant eigenfunctions and their associated eigenenergies,
one can calculate the regularized scalar and pseudoscalar
densities $S (r)$ and $P (r)$.
With use of these $S(r)$ and $P(r)$, Eq.(\ref{eq:profile})
can then be used to obtain a new soliton profile 
$F (r)$. The whole procedure above is repeated with this new profile 
$F (r)$ until the self-consistency is attained.

\section{Generalized Form Factors in the CQSM}

 Since the generalized form factors of the nucleon are given as moments
of generalized parton distributions (GPDs), it is convenient to start
with the theoretical expressions of the unpolarized GPDs
$H (x, \xi, t)$ and $E (x, \xi, t)$ within the CQSM.
Following the notation in \cite{Ossmann05},\cite{WT05},
we introduce the quantities
\begin{equation}
 {\cal M}_{s^{\prime} s}^{(I = 0)} \equiv \int \frac{d \lambda}{2 \pi} \,
 e^{i \lambda x} \,\langle \vecvar{p}^{\prime}, s^{\prime} 
 | \,\bar{\psi} \left(-\frac{\lambda n}{2} \right) \not\!n 
 \psi \left(\frac{\lambda n}{2} \right)
 | \vecvar{p}, s \rangle ,
\end{equation}
and 
\begin{equation}
 {\cal M}_{s^{\prime} s}^{(I = 1)} \equiv \int \frac{d \lambda}{2 \pi} \,
 e^{i \lambda x} \,\langle \vecvar{p}^{\prime}, s^{\prime} 
 | \,\bar{\psi} \left(-\frac{\lambda n}{2} \right) \tau_3 \not\!n 
 \psi \left(\frac{\lambda n}{2} \right) | \vecvar{p}, s \rangle .
\end{equation}
Here, the isoscalar and isovector combinations respectively
correspond to the sum and the difference of the quark flavors
$u$ and $d$.
The relation between these quantities and the generalized parton
distribution functions $H(x, \xi, t)$ and $E(x, \xi, t)$
are obtained most conveniently 
in the so-called Breit frame. They are given by
\begin{eqnarray}
 {\cal M}_{s^{\prime} s}^{(I = 0)} 
 &=& 2 \,\delta_{s^{\prime} s} \,H_E^{(I = 0)} (x, \xi, t)
 \ - \ \frac{i \,\varepsilon^{3 k l} \,\Delta^k}{M_N} \,
 (\sigma^l)_{s^{\prime} s} \, E_M^{(I = 0)} (x, \xi, t) ,\\
 {\cal M}_{s^{\prime} s}^{(I = 1)} 
 &=& 2 \,\delta_{s^{\prime} s} \,H_E^{(I = 1)} (x, \xi, t)
 \ - \ \frac{i \,\varepsilon^{3 k l} \,\Delta^k}{M_N} \,
 (\sigma^l)_{s^{\prime} s} \, E_M^{(I = 1)} (x, \xi, t) .
\end{eqnarray}
where
\begin{eqnarray}
 H_E^{(I = 0/1)} (x, \xi, t) &\equiv& H^{(I = 0/1)} (x, \xi, t)
 + \frac{t}{4 M_N^2} E^{(I = 0/1)} (x, \xi, t) ,\\
 E_M^{(I = 0/1)} (x, \xi, t) &\equiv& H^{(I = 0/1)} (x, \xi, t)
 + E^{(I = 0/1)} (x, \xi, t) .
\end{eqnarray}
These two independent combinations of $H (x, \xi, t)$ and $E (x, \xi, t)$
can be extracted through the spin projection of ${\cal M}^{(I = 0/1)}$ as
\begin{eqnarray}
 H_E^{(I)} (x, \xi, t) &=& \frac{1}{4} \,
 \mbox{tr} \{ {\cal M}^{(I)} \} , \label{eq:HEtrace} \\
 E_M^{(I)} (x, \xi, t) &=& 
 \frac{i \,M_N \,\varepsilon^{3 b m} \,
 \Delta^b}{2 \vecvar{\Delta}_{\perp}^2}
 \,\mbox{tr} \{ \sigma^m {\cal M}^{(I)} \} , \label{eq:EMtrace}
\end{eqnarray}
where ``tr" denotes the trace over spin indices, while 
$\vecvar{\Delta}_{\perp}^2 = \vecvar{\Delta}^2 - (\Delta^3)^2
 = -t - (-2 M_N \xi)^2$. Now, within the CQSM, it is possible to evaluate
the right-hand side (rhs) of (\ref{eq:HEtrace}) and (\ref{eq:EMtrace}).
Since the answers are already given in several previous
papers \cite{Ossmann05}--\cite{Petrov98}, 
we do not repeat the derivation. Here we comment only on the following
general structure of the theoretical expressions for relevant
observables in the CQSM. The leading contribution 
just corresponds to the mean field prediction, which is independent of 
the collective rotational velocity $\Omega$ of the hedgehog soliton.  
The next-to-leading order term takes account of the linear response of 
the internal quark motion to the rotational motion as an external 
perturbation, and consequently it is proportional to $\Omega$.
It is known that the leading-order term contributes to the isoscalar 
combination of $H_E (x, \xi, t)$ and to the isovector combination of 
$E_M (x, \xi, t)$, while the isoscalar part of $H_E (x, \xi, t)$ and the
isovector part of $E_M (x, \xi, t)$ survived only at the next-to-leading 
order of $\Omega$ (or of $1/N_c$). The leading-order GPDs are then
given as
\begin{eqnarray}
 H_E^{(I = 0)} (x, \xi, t) &=& M_N N_c \int \frac{d z^0}{2 \pi} \,
 \sum_{n \leq 0} \,e^{i z^0 (x M_N - E_n)} \int d^3 \vecvar{x} \,
 \Phi_n^{\dagger} (\vecvar{x}) \nonumber \\
 &\,& \times \ (1 + \gamma^0 \gamma^3) \,e^{-i(z^0/2) \hat{p}_3} \,
 e^{i \vecvar{\Delta} \cdot \vecvar{x}} \,e^{-i (z^0/2) \hat{p}_3} \,
 \Phi_n (\vecvar{x}) , \label{eq:GPDHEis} \\
 E_M^{(I = 1)} (x, \xi, t) 
 &=& \frac{2 \,i \,M_N^2 \,N_c}{3 \,(\vecvar{\Delta}_{\perp})^2} 
 \int \frac{d z^0}{2 \pi} \,\sum_{n \leq 0} \,e^{i z^0 (x M_N - E_n)} \,
 \int d^3 \vecvar{x} \,\Phi_n^{\dagger} (\vecvar{x}) \nonumber \\
 &\,& \times \ (1 + \gamma^0 \gamma^3) \,
 (\vecvar{\tau} \times \vecvar{\Delta})^3 \,
 e^{-i(z^0/2) \hat{p}_3} \,e^{i \vecvar{\Delta} \cdot \vecvar{x}} 
 \,e^{-i (z^0/2) \hat{p}_3} \,\Phi_n (\vecvar{x}) . \label{eq:GPDEMis}
\end{eqnarray}
Here the symbol $\Sigma_{n \leq 0}$ denotes the summation over the
occupied (the valence plus negative-energy Dirac sea) quark
orbitals in the hedgehog mean field.
On the other hand, the theoretical expressions for the isovector 
part of $H_E$ and the isoscalar part of $E_M$, which
survive at the next-to-leading order, are a little more complicated.
They are given as double sums over the single quark orbitals as
\begin{eqnarray}
 &\,& H_E^{(I = 1)} (x, \xi, t) \nonumber \\
 &=& - \frac{M_N N_c}{12 I} \int \frac{d z^0}{2 \pi} \,
 \left[\, \left\{ \sum_{m = all, n \leq 0 (E_m \neq E_n)} \,
 e^{-i E_n z^0} \ - \ 
 \sum_{n = all, m \leq 0 (E_m \neq E_n)} \,e^{-i E_m z^0} \right\} 
 \frac{1}{E_n - E_m} \right. \nonumber \\
 &\,& \left. \hspace{50mm} + \  \frac{1}{M_N} \,\frac{d}{d x} \,
 \sum_{m = all, n \leq 0 (E_m \neq E_n)} \,e^{-i E_n z^0} \right] \,
 e^{i x M_N z^0} \nonumber \\
 &\,& \hspace{30mm} \times \ \langle n | \tau^a | m \rangle 
 \langle m | \,\tau^a \,(1 + \gamma^0 \gamma^3) \,
 e^{-i (z^0/2) \hat{p}_3} \,
 e^{i \vecvar{\Delta} \cdot \vecvar{x}} \,e^{-i (z^0/2) \hat{p}_3}
 | n \rangle . \label{eq:GPDHEiv}
\end{eqnarray}
and
\begin{eqnarray}
 &\,& E_M^{(I = 0)} (x, \xi, t) \nonumber\\
 &=& i \,\frac{M_N^2 N_c}{2 I} \int \frac{d z^0}{2 \pi} \,
 \left[\, \left\{ \sum_{m = all, n \leq 0 (E_m \neq E_n)} \,
 e^{-i E_n z^0} \ - \ 
 \sum_{n = all, m \leq 0 (E_m \neq E_n)} e^{-i E_m z^0} \right\} 
 \frac{1}{E_n - E_m} \right. \nonumber \\
 &\,& \left. \hspace{50mm} + \frac{1}{M_N} \,\frac{d}{d x} 
 \sum_{m = all, n \leq 0 (E_m \neq E_n)} e^{-i E_n z^0} \right]
 e^{i x M_N z^0} \nonumber \\
 &\,& \hspace{25mm} \times \ \langle n | \tau^b | m \rangle 
 \langle m | \,(1 + \gamma^0 \gamma^3) \,e^{-i (z^0/2) \hat{p}_3} \,
 \frac{\varepsilon^{3 a b} \Delta^a}{\vecvar{\Delta}_{\perp}^2} \,
 e^{i \vecvar{\Delta} \cdot \vecvar{x}} \,e^{-i (z^0/2) \hat{p}_3}
 | n \rangle  . \label{eq:GPDEMiv}
\end{eqnarray}
These four expressions for the unpolarized GPDs,
i.e. (\ref{eq:GPDHEis}) $\sim$ (\ref{eq:GPDEMiv}),
are the basic starting equations for our present study 
of the generalized form factors of the nucleon within the CQSM.
There are infinite tower of generalized form factors, which are defined
as the $n$-th moments of GPDs.
In the present study, we confine ourselves to the 
1st and the 2nd moments, which respectively corresponds to the standard 
electromagnetic form factors of the nucleon and the so-called
gravitational form factors.
We are especially interested in the second one, since they are believed
to contain valuable information on the spin contents of the nucleon
through Ji's angular momentum sum rule \cite{Ji97},\cite{HJL99}.
For each isospin channel, the 1st and the 2nd moments of
$H_E^{(I)} (x, 0, t)$ define the Sachs-electric and gravito-electric
form factors as 
\begin{equation}
 G_{E, 10}^{(I = 0/1)} (t) \equiv \int_{-1}^1 \,
 H_E^{(I =0/1)} (x, 0, t) \,d x , \label{eq:GE10def}
\end{equation}
and 
\begin{equation}
 G_{E, 20}^{(I = 0/1)} (t) \equiv \int_{-1}^1 \,x \,
 H_E^{(I =0/1)} (x, 0, t) \,d x .
\end{equation}
On the other hand, the 1st and the 2nd moments of $E_M^{(I)} (x, 0, t)$
respectively define the Sachs-magnetic and gravito-magnetic form factors
as
\begin{equation}
 G_{M, 10}^{(I = 0/1)} (t) \equiv \int_{-1}^1 \,
 E_M^{(I = 0/1)} (x, 0, t) \,d x ,
\end{equation}
and
\begin{equation}
 G_{M, 20}^{(I = 0/1)} (t) \equiv \int_{-1}^1 \,
 x \,E_M^{(I = 0/1)} (x, 0, t) \,d x . \label{eq:GM2nd}
\end{equation}
In the following, we shall explain how we can calculate the generalized
form factors based on the theoretical expressions of corresponding GPDs,
by taking $G_{E, 10}^{(I = 0)} (t)$ and $G_{E, 20}^{(I = 0)} (t)$
as examples. Setting $\xi = 0$ and integrating over $z^0$ in
(\ref{eq:GPDHEis}), we obtain
\begin{eqnarray}
 H_E^{(I = 0)} (x, 0, t) &=& M_N \,N_c \, \int d^3 x \,
 e^{i \vecvar{\Delta}_{\perp} \cdot \vecvar{x}} \nonumber \\
 &\times& \sum_{n \leq 0} \,\Phi_n^{\dagger} (\vecvar{x}) \,
 (1 + \gamma^0 \gamma^3) \,\delta (x M_N - E_n - \hat{p}_3) \,
 \Phi_n (\vecvar{x}) . \label{eq:HEis}
\end{eqnarray}
Putting this expression into (\ref{eq:GE10def}), we have
\begin{eqnarray}
 G_{E, 10}^{(I = 0)} (t) &=& N_c \int d^3 x \,
 e^{i \vecvar{\Delta}_{\perp} \cdot \vecvar{x}} \,
 \sum_{n \leq 0} \Phi_n^{\dagger} (\vecvar{x})
 (1 + \gamma^0 \gamma^3) \Phi_n (\vecvar{x}) . 
\end{eqnarray}
It is easy to see that, using the generalized spherical symmetry of the 
hedgehog configuration, the term containing the factor
$\gamma^0 \gamma^3$ identically vanishes, so that $G_{E,10}^{(I=0)}(t)$
is reduced to a simple form as follows :
\begin{equation}
 G_{E, 10}^{(I = 0)} (t) = N_c \int d^3 x 
 \sum_{n \leq 0} \Phi_n^{\dagger} (\vecvar{x}) \,j_0 (\Delta_{\perp} x)
 \,\Phi_n (\vecvar{x}) .
\end{equation}
Aside from the factor $N_c \,( = 3)$, this is 
nothing but the known expression for the isoscalar Sachs-electric
form factor of the nucleon in the CQSM \cite{Christov96}.

A less trivial example is $G_{E, 20}^{(I = 0)} (t)$, which is defined
as the 2nd moment of  $H_E^{(I = 0)} (x, 0, t)$.
Inserting (\ref{eq:HEis}) into (\ref{eq:GE10def}) and carrying out the 
integration over $x$, we obtain
\begin{eqnarray}
 G_{E, 20}^{(I = 0)} (t) &=& \frac{1}{M_N} \,N_c \,\int d^3 x
 \,e^{i \vecvar{\Delta}_{\perp} \cdot \vecvar{x}}
 \, \sum_{n \leq 0} \,\Phi_n^{\dagger} (\vecvar{x}) \,
 (1 + \gamma^0 \gamma^3) \,(E_n + \hat{p}_3) \,\Phi_n (\vecvar{x}) .
\end{eqnarray}
Using the partial-wave expansion of 
$e^{i \vecvar{\Delta}_{\perp} \cdot \vecvar{x}}$, this can be written as
\begin{eqnarray}
 G_{E, 20}^{(I = 0)} (t) &=& \frac{1}{M_N} \,N_c \,\int d^3 x \,
 \sum_{l, m} \,4 \pi \,i^l \,
 Y_{l m}^* (\hat{\Delta}_{\perp}) \nonumber \\
 &\times& \sum_{n \leq 0} \Phi_n^{\dagger} (\vecvar{x}) \,
 j_l (\Delta_{\perp} x) \,Y_{l m}(\hat{x}) \,(1 + \alpha_3) \,
 (E_n + \hat{p}_3) \,\Phi_n (\vecvar{x}) .
\end{eqnarray}
This can further be divided into four pieces as
\begin{equation}
 G_{E, 20}^{(I = 0)} (t) = \sum_{i = 1}^4 G_i ,
\end{equation}
where
\begin{equation}
 G_i = \frac{1}{M_N} \,N_c \,\int d^3 x \,\sum_{l, m} \,4 \pi \,i^l \,
 Y_{l m}^* (\hat{\Delta}_{\perp}) \,M_i ,
\end{equation}
with
\begin{eqnarray}
 M_1 &=& \sum_{n \leq 0} \,\Phi_n^{\dagger} (\vecvar{x}) \,
 j_l (\Delta_{\perp} x) \,Y_{l m} (\hat{x}) \,E_n \,
 \Phi_n (\vecvar{x}) , \\
 M_2 &=& \sum_{n \leq 0} \,\Phi_n^{\dagger} (\vecvar{x}) \,
 j_l (\Delta_{\perp} x) \,Y_{l m} (\hat{x}) \,\alpha_3 E_n \,
 \Phi_n (\vecvar{x}) , \\
 M_3 &=& \sum_{n \leq 0} \,\Phi_n^{\dagger} (\vecvar{x}) \,
 j_l (\Delta_{\perp} x) \,Y_{l m} (\hat{x}) \,\hat{p}_3 \,
 \Phi_n (\vecvar{x}) , \\
 M_4 &=& \sum_{n \leq 0} \,\Phi_n^{\dagger} (\vecvar{x}) \,
 j_l (\Delta_{\perp} x) \,Y_{l m} (\hat{x}) \,\alpha_3 \,\hat{p}_3 \,
 \Phi_n (\vecvar{x}) . \label{eq:M4}
\end{eqnarray}
To proceed further, we first notice that, by using the generalized
spherical symmetry, $M_1$ survives only when $l = m = 0$, i.e. 
\begin{equation}
 M_1 \ \sim \ \sum_{n \leq 0} \,\Phi_n^{\dagger} (\vecvar{x}) \,
 j_0 (\Delta_{\perp} x) \,\frac{\delta_{l,0} \delta_{m,0}}{\sqrt{4 \pi}}
 \,E_n \Phi_n (\vecvar{x}) ,
\end{equation}
which leads to the result :
\begin{equation}
 G_1 = \frac{N_c}{M_N} \,\int \,d^3 x \,\sum_{n \leq 0} \,
 \Phi_n^{\dagger} (\vecvar{x}) \,
 j_0 (\Delta_{\perp} x) \,E_n \,\Phi_n (\vecvar{x}) .
\end{equation}
To evaluate $G_2$, we first note that 
\begin{equation}
 Y_{l m} (\hat{x})  \times \alpha_3 
 = \sum_{\lambda} \,\langle l m 1 0 | \lambda m \rangle \,
 [ Y_l (\hat{x}) \times \vecvar{\alpha} ]^{(\lambda)} .
\end{equation}
Here, the generalized spherical symmetry dictates that $\lambda$ must
be zero, so that the rhs of the above equation is effectively reduced to
\begin{equation}
 - \,\frac{1}{\sqrt{3}} \,\delta_{l,1} \, \delta_{m,0}
 [ Y_1 (\hat{x}) \times \vecvar{\alpha} ]^{(0)} .
\end{equation}
This then gives
\begin{eqnarray}
 G_2 &=& \frac{1}{M_N} \,N_c \,\int \,d^3 x \,4 \pi \,i \,
 Y_{10}^* (\hat{\Delta}_{\perp}) \,\left(-\frac{1}{\sqrt{3}} \right) \,
 \,\sum_{n \leq 0} \Phi_n^{\dagger} (\vecvar{x})
 j_1 (\Delta_{\perp} x) \,
 [ Y_1 (\hat{x}) \times \vecvar{\alpha} ]^{(0)} \,
 \Phi_n (\vecvar{x}) .
\end{eqnarray}
Owing to the identity
\begin{equation}
 Y_{10} (\hat{\Delta}_{\perp}) 
 \ = \ \sqrt{\frac{3}{4 \pi}} \,P_1 (\cos \frac{\pi}{2}) \ = \ 0 ,
\end{equation}
we therefore find that
\begin{equation}
 G_2 \ = \ 0 .
\end{equation}
Next we investigate the third term $G_3$. Using
\begin{eqnarray}
 Y_{l m } (\hat{x}) \times \hat{p}_3 &=& \sum_{\lambda} \,
 \langle l m 1 0 | \lambda m \rangle \,
 [ Y_l (\hat{x}) \times \hat{\vecvar{p}} ]^{(\lambda)}
 \ \sim \ - \,\frac{1}{\sqrt{3}} \,\delta_{l,1} \,\delta_{m,0} \,
 [ Y_1 (\hat{x}) \times \hat{\vecvar{p}} ]^{(0)} ,
\end{eqnarray}
 we obtain 
\begin{eqnarray}
 G_3 &=& \frac{1}{M_N} \,N_c \,\int d^3 x \,4 \pi \,i \,
 Y_{10}^* (\hat{\Delta}_{\perp}) \,\left(- \,\frac{1}{\sqrt{3}} \right)
 \, \sum_{n \leq 0} \,\Phi_n^{\dagger} (\vecvar{x}) \,
 j_1 (\Delta_{\perp} x) \,
 [ Y_1 (\hat{x}) \times \hat{\vecvar{p}} ]^{(0)} \,
 \Phi_n (\vecvar{x}) .
\end{eqnarray}
This term vanishes by the same reason as $G_2$ does. The last term $G_4$ 
is a little more complicated.  We first notice that 
\begin{eqnarray}
 Y_{l m} (\hat{x}) \,\alpha_3 \,\hat{p}_3
 &=& Y_{l m} (\hat{x}) \,\sum_{\lambda} \,
 \langle 10 10 | \lambda 0 \rangle \,
 [ \vecvar{\alpha} \times \hat{\vecvar{p}} ]^{(\lambda)}_0 \nonumber \\
 &\sim& \sum_{\lambda} \,
 \langle 1 0 1 0 | \lambda 0 \rangle \,
 \langle l m \lambda 0 | 00 \rangle \,
 [ Y_l (\hat{x}) \times 
 [ \vecvar{\alpha} \times \hat{\vecvar{p}} ]^{(\lambda)} ]^{(0)} 
 \nonumber \\
 &=& \delta_{m, 0} \,\langle 1 0 1 0 | l 0 \rangle \,
 \langle l 0 l 0 | 0 0 \rangle \,
 [ Y_l (\hat{x}) \times 
 [ \vecvar{
 \alpha} \times \hat{\vecvar{p}} ]^{(l)} ] ^{(0)} ,
\end{eqnarray}
which dictates that $l$ must be 0 or 2. 
Inserting the above expression into (\ref{eq:M4}),
and using the explicit values of Clebsch-Gordan coefficients, $G_4$ 
becomes
\begin{eqnarray}
 G_4 &=& -\frac{1}{\sqrt{3}} \cdot \frac{N_c}{M_N}
 \int d^3 x \,\sum_{n \leq 0} \,\Phi_n^{\dagger} (\vecvar{x}) \,
 j_0 (\Delta_{\perp} x) \,
 [ \vecvar{\alpha} \times \hat{\vecvar{p}} ]^{(0)} \,
 \Phi_n (\vecvar{x}) \nonumber \\
 &+& \frac{\sqrt{4 \pi}}{\sqrt{6}} \cdot \frac{N_c}{M_N} 
 \int d^3 x \,\sum_{n \leq 0} \,\Phi_n^{\dagger} (\vecvar{x}) \,
 j_2 (\Delta_{\perp} x) \,
 [ Y_2 (\hat{x}) \times 
 [ \vecvar{\alpha} \times \hat{\vecvar{p}} ]^{(2)} ]^{(0)} \,
 \Phi_n (\vecvar{x}) .
\end{eqnarray}
Using the identities
\begin{eqnarray}
 \,[ \vecvar{\alpha} \times \hat{\vecvar{p}} ]^{(0)}
 &=& -\frac{1}{\sqrt{3}} \vecvar{\alpha} \cdot \hat{\vecvar{p}} , \\
 \,[ Y_2 (\hat{x}) 
 \times [ \vecvar{\alpha} \times \vecvar{p} ]^{(2)} ]^{(0)}
 &=& [ [ Y_2 (\hat{x}) \times \hat{\vecvar{p}} ]^{(1)} 
 \times \vecvar{\alpha} ]^{(0)} ,
\end{eqnarray}
$G_4$ can also be written as 
\begin{eqnarray}
 G_4 &=& \frac{1}{3} \cdot \frac{N_c}{M_N} \,
 \int \,d^3 x \,\sum_{n \leq 0} \,\Phi_n^{\dagger} (\vecvar{x}) \,
 j_0 (\Delta_{\perp} x) \,
 \vecvar{\alpha} \cdot \hat{\vecvar{p}} \,
 \Phi_n (\vecvar{x}) \nonumber \\
 &+& \frac{\sqrt{4 \pi}}{\sqrt{6}} \cdot \frac{N_c}{M_N} 
 \int \,d^3 x \,\sum_{n \leq 0} \,\Phi_n^{\dagger} (\vecvar{x}) \,
 j_2 (\Delta_{\perp} x) \,
 [ Y_2 (\hat{x}) \times \hat{\vecvar{p}} ]^{(1)}
 \times \vecvar{\alpha} ]^{(0)} \,
 \Phi_n (\vecvar{x}) .
\end{eqnarray}
Collecting the answers for $G_1, G_2, G_3$ and $G_4$, we finally obtain
\begin{eqnarray}
 G_{E, 20}^{(I = 0)} (t)
 &=& \frac{N_c}{M_N} \int \,d^3 x \,
 \sum_{n \leq 0} \,\Phi_n^{\dagger} (\vecvar{x}) \,
 j_0 (\Delta_{\perp} x) \,E_n \,\Phi_n (\vecvar{x}) \nonumber \\
 &+& \frac{1}{3} \cdot \frac{N_c}{M_N} \,\int \,d^3 x \,
 \sum_{n \leq 0} \,\Phi_n^{\dagger} (\vecvar{x}) \,
 j_0 (\Delta_{\perp} x) \,\vecvar{\alpha} \cdot \hat{\vecvar{p}} \,
 \Phi_n (\vecvar{x}) \nonumber \\
 &+& \frac{\sqrt{4 \pi}}{\sqrt{6}} \cdot \frac{N_c}{M_N} \,
 \int \,d^3 x \,
 \sum_{n \leq 0} \,\Phi_n^{\dagger} (\vecvar{x}) \,
 j_2 (\Delta_{\perp} x) \,
 [ Y_2 (\hat{x}) \times \hat{\vecvar{p}} ]^{(1)} 
 \times \vecvar{\alpha} ]^{(0)} \,\Phi_n (\vecvar{x}) .
\end{eqnarray}

Up to now, we have obtained the theoretical expressions for the
isoscalar combination of the generalized form factors
$G_{E, 10}^{(I = 0)} (t)$ and 
$G_{E, 20}^{(I = 0)} (t)$. For notational convenience, we summarize
these results in a little more compact forms as follows :
\begin{eqnarray}
 G_{E, 10}^{(I = 0)} (t) &=& \int_{-1}^1 
 H_E^{(I = 0)} (x, 0, t) d x
 \ = \ N_c \,\sum_{n \leq 0} \,
 \langle n | j_0 (\Delta_{\perp} r) | n \rangle , \label{eq:IntHE10is}
\end{eqnarray}
and 
\begin{eqnarray}
 G_{E, 20}^{(I = 0)} (t) &=& \int_{-1}^1 \,x 
 H_E^{(I = 0)} (x, 0, t) \,d x \nonumber \\
 &=& \frac{1}{M_N} \,\left\{ N_c \,\sum_{n \leq 0} \,E_n \,
 \langle n | j_0 (\Delta_{\perp} r) | n \rangle \right. \nonumber \\
 &+& N_c \,\sum_{n \leq 0} \,
 \langle n | j_0 (\Delta_{\perp} r) \,\frac{1}{3} \,
 \vecvar{\alpha} \cdot \vecvar{p} \,| n \rangle \nonumber \\
 &+& \left. \frac{\sqrt{4 \pi}}{\sqrt{6}} \,N_c \,
 \sum_{n \leq 0} \,
 \langle n | j_2 (\Delta_{\perp} r) \,[ [Y_2 (\hat{r}) 
 \times \vecvar{p} ]^{(1)} \times \vecvar{\alpha} ]^{(0)} \,
 | n \rangle \right\} . \label{eq:IntHE20is}
\end{eqnarray}
As pointed out before, $G_{E, 10}^{(I = 0)} (t)$ is $N_c \,( = 3)$
times the isoscalar combination of standard Sachs-electric form factor
of the nucleon.
Analogously, we may call $G_{E, 20}^{(I = 0)} (t)$ the
gravitoelectric form factor of the nucleon (its quark part),
since it is related to the nonforward nucleon matrix elements of
the quark part of the QCD energy momentum tensor.

The other generalized form factors can be obtained in a similar way.
The isovector part of the generalized electric form factors survive only 
at the next-to-leading order of $\Omega$. They are given as 
\begin{eqnarray}
 G_{E, 10}^{(I = 1)} (t) &=& \int_{-1}^1 H_E^{(I = 1)} (x, 0, t) \,
 d x \nonumber \\
 &=& \frac{1}{3 I} \,\left(\frac{N_c}{2} \right) \,\sum_{m > 0, n \leq 0}
 \frac{1}{E_m - E_n} \,\langle m || \vecvar{\tau} || n \rangle
 \langle m || j_0 (\Delta_{\perp} r) \vecvar{\tau} || n \rangle ,
 \label{eq:IntHE10iv}
\end{eqnarray}
and 
\begin{eqnarray}
 G_{E, 20}^{(I = 1)} (t) &=& 
 \int_{-1}^1 \,x \,H_E^{(I = 1)} (x, 0, t) \,
 d x \nonumber \\
 &=& \frac{1}{M_N} \cdot \frac{1}{3 I} \,\left(\frac{N_c}{2} \right) \,
 \sum_{m > 0, n \leq 0} \,
 \frac{1}{E_m - E_n} \,\langle m || \vecvar{\tau} || n \rangle 
 \nonumber \\
 &\times& \left\{ \,\frac{E_m + E_n}{2} \,
 \langle m || j_0 (\Delta_{\perp} r) \vecvar{\tau} || n \rangle \right.
 \nonumber \\
 &+& \langle m || \,j_0 (\Delta_{\perp} r) \,\frac{1}{3} \,
 (\vecvar{\alpha} \cdot \vecvar{p}) \,\vecvar{\tau} \,|| n \rangle
 \nonumber \\
 &+& \left. \frac{\sqrt{4 \pi}}{\sqrt{6}} \,
 \langle m || \,j_2 (\Delta_{\perp} r) \,
 [ [Y_2 (\hat{r}) \times \vecvar{p} ]^{(1)} \times 
 \vecvar{\alpha} ]^{(0)} \,\vecvar{\tau} \,|| n \rangle \right\} .
\end{eqnarray}
The isoscalar combination of the generalized magnetic form factors also 
survive only at the next-to-leading order of $\Omega$, so that they are 
given as double sums over the single-quark orbitals in the hedgehog mean 
field as 
\begin{eqnarray}
 G_{M, 10}^{(I = 0)} (t) &=& \int_{-1}^1 \,
 E_M^{(I = 0)} (x, 0, t) \,d x \nonumber \\
 &=& - \,\frac{M_N}{I} \,\left(\frac{N_c}{2} \right) \,
 \sum_{m > 0, n \leq 0} \,
 \frac{1}{E_m - E_n} \,\langle m || \vecvar{\tau} || n \rangle \,
 \langle m || \,\frac{j_1 (\Delta_{\perp} r)}{\Delta_{\perp} r} \,
 (\vecvar{r} \times \vecvar{\alpha}) \,|| n \rangle ,
\end{eqnarray}
and
\begin{eqnarray}
 G_{M, 20}^{(I = 0)} (t)
 &=& \int_{-1}^1 \,x E_M^{(I = 0)} (x, 0, t) \,d x \nonumber \\
 &=& - \,\frac{1}{I} \,\left(\frac{N_c}{2} \right) \,
 \sum_{m > 0, n \leq 0} \,\frac{1}{E_m - E_n} \,
 \langle m || \vecvar{\tau} || n \rangle \nonumber \\
 &\times& \left\{\, \frac{E_m + E_n}{2} \,
 \langle m || \,\frac{j_1 (\Delta_{\perp} r)}{\Delta_{\perp} r} \,
 (\vecvar{r} \times \vecvar{\alpha}) \,|| n \rangle \ + \ 
 \langle m || \,\frac{j_1 (\Delta_{\perp} r)}{\Delta_{\perp} r} \,
 \vecvar{L} \,|| n \rangle \right\} .
\end{eqnarray}
We recall that $G_{M, 10}^{(I = 0)} (t)$ just coincides with the known 
expression of the isoscalar Sachs-magnetic form factor of the nucleon
in the CQSM \cite{Christov95}.
On the other hand, $G_{M, 20}^{(I = 0)} (t)$ is sometimes called the 
gravitomagnetic form factor of the nucleon (its isoscalar part), which 
we can evaluate within the QCSM based on the above theoretical
expression. Finally, the leading-order contribution to the isovector
part of the generalized magnetic form factors are given as 
\begin{eqnarray}
 G_{M, 10}^{(I = 1)} (t) 
 &=& \int_{-1}^1  E_M^{(I = 1)} (x, 0, t) \,d x
 \ = \ - \,\frac{M_N}{3} \cdot N_c \,\sum_{n \leq 0} \,
 \langle n || \,\frac{j_1 (\Delta_{\perp} r)}{\Delta_{\perp} r} \,
 \vecvar{\tau} \cdot (\vecvar{r} \times \vecvar{\alpha})\,|| n \rangle ,
 \ \ \ \label{eq:MGivform}
\end{eqnarray}
and
\begin{eqnarray}
 G_{M, 20}^{(I = 1)} (t) 
 &=& \int_{-1}^1 x E_M^{(I = 1)} (x, 0, t) \,d x \nonumber \\
 &=& - \frac{1}{3} \cdot N_c \,\sum_{n \leq 0} \,
 \left\{\, E_n \,\langle n || \,
 \frac{j_1 (\Delta_{\perp} r)}{\Delta_{\perp} r} \,
 \vecvar{\tau} \cdot (\vecvar{r} \times \vecvar{\alpha})\,|| n \rangle 
 \ + \ \langle n || \,\frac{j_1 (\Delta_{\perp} r)}{\Delta_{\perp} r} 
 \vecvar{\tau} \cdot \vecvar{L} \,|| n \rangle \right\}. 
 \label{eq:MG2ivform} \ \ \ \ \ 
\end{eqnarray}
Especially interesting to us are the values of the generalized form 
factors in the forward limit $t \rightarrow 0$.
The consideration of this 
limit is also useful for verifying consistency of our theoretical
analyses, since it leads to fundamental sum rules discussed below.
We first consider the forward limit of $G_{E, 10}^{(I = 0)} (t)$.
From (\ref{eq:IntHE10is}), we find that
\begin{eqnarray}
 G_{E, 10}^{(I = 0)} (t = 0) &=& \int_{-1}^1 
 H_E^{(I = 0)} (x, 0, 0) \,d x
 \ = \ N_c \,\sum_{n \leq 0} \,1 .
\end{eqnarray}
Subtracting the corresponding vacuum contribution, this reduces to 
$N_c \,(= 3)$. If we remember the relation 
\begin{eqnarray}
 \int_{-1}^1 H_E^{(I = 0)} (x, 0, 0) d x 
 &=& \int_{-1}^1 H^{u + d} (x, 0, 0) d x
 \ = \ \int_{-1}^1 f^{u+ d} (x) d x 
 \ = \ N^u + N^d ,
\end{eqnarray}
the forward limit of (\ref{eq:IntHE10is}) just leads to the sum rule :
\begin{equation}
 G_{E, 10}^{(I = 0)} (t = 0) = N^u + N^d = 3 , \label{eq:udpSR}
\end{equation}
which denotes that the sum of the $u$-quark and $d$-quark numbers
in the proton is three.

Next we turn to the forward limit of $G_{E, 20}^{(I = 0)} (t)$,
which gives 
\begin{eqnarray}
 G_{E, 20}^{(I = 0)} (t = 0) &=& \frac{1}{M_N} \,
 \left\{\, N_c \,
 \sum_{n \leq 0} \,E_n \ + \ 
 \frac{1}{3} \,N_c \,\sum_{n \leq 0} \,
 \langle n | \vecvar{\alpha} \cdot \vecvar{p} | n \rangle \right\} .
 \label{eq:MomSum}
\end{eqnarray}
It is easy to see that, after regularization and vacuum subtraction,
the first term of the rhs of the 
above equation reduces to the fermion (quark) part of the soliton energy,
i.e. $E_F^{reg}$ in (\ref{eq:Estatic}).
It was proved in \cite{DPPPW96} that, in the CQSM 
with vanishing pion mass, the following identity holds :
\begin{equation}
 \sum_{n \leq 0} \,
 \langle n | \vecvar{\alpha} \cdot \vecvar{p} | n \rangle
 = 0
\end{equation}
In the case of finite pion mass, which we are handling, this identity
does not hold. Instead, we can prove (see Appendix) that
\begin{equation}
 \frac{1}{3} \,N_c \,\sum_{n \leq 0} \,
 \langle n | \vecvar{\alpha} \cdot \vecvar{p} | n \rangle = E_M .
 \label{eq:alfp}
\end{equation}
That is, the second term in the parenthesis of rhs of
eq.(\ref{eq:MomSum}) just coincides with the pion part of the soliton
energy (or mass).
Since the sum of the quark and pion part give the total soliton mass
$M_N$, we then find that
\begin{equation}
 G_{E, 20}^{(I = 0)} (t = 0) 
 = \frac{1}{M_N} \cdot M_N = 1 .
\end{equation}
In consideration of eq.(\ref{eq:IntHE20is}), this relation can also
be expressed as
\begin{eqnarray}
 \int_{-1}^1 x H_E^{(I = 0)} (x, 0, 0) \,d x 
 &=& \int_{-1}^1 x H^{u + d} (x, 0, 0) \,d x
 \ = \ \int_{-1}^1 \,x f^{u + d} (x) \,dx \ = \ 
 \langle x \rangle^{u + d} = 1 , \ \ \ \label{eq:momSR}
\end{eqnarray}
which means that the total momentum fraction carried by quark fields 
(the $u$- and $d$-quarks) is just unity. 
This is an expected result, since the CQSM contains quark fields only
(note that the pion is not an independent field of quarks), so that the
total nucleon momentum should be saturated by the quark fields 
alone.

Taking the forward limit of $G_{E, 10}^{(I = 1)} (t)$, we are again
led to a trivial sum rule, constrained by the conservation low.
In fact, we have
\begin{eqnarray}
 G_{E, 10}^{(I = 1)} (t= 0) &=& \frac{1}{I} \,
 \left(\frac{N_c}{6} \right)
 \sum_{m > 0, n \leq 0} \,\frac{1}{E_m - E_n} \,
 \langle m || \vecvar{\tau} || n \rangle^2
 \ = \ \frac{1}{I} \cdot I = 1 ,
\end{eqnarray}
thereby leading to
\begin{eqnarray}
 \int_{-1}^1 H_E^{(I = 1)} (x, 0, 0) \,d x 
 &=& \int_{-1}^1 H^{u-d} (x, 0, 0) \,d x
 \ = \ \int_{-1}^1 f^{u - d} (x) \,d x 
 \ = \ N^u - N^d = 1 , \ \ \ \label{eq:udmSR}
\end{eqnarray}
which denotes that the difference of the $u$-quark and the
$d$-quark numbers in the proton is just unity.
On the other hand, the forward
limit of $G_{E, 20}^{(I = 1)} (t)$ leads to the first nontrivial sum
rule as
\begin{eqnarray}
 G_{E, 20}^{(I = 1)} (t = 0) 
 &=& \int_{-1}^1 \,x H_E^{(I = 1)} (x, 0, 0) \,d x
 \ = \ \int_{-1}^1 \,x f^{u - d} (x) \,d x
 \ = \ \langle x \rangle^{u - d} \nonumber \\
 &=& \frac{1}{M_N} \,\frac{1}{I} \,
 \left(\frac{N_c}{6} \right) \,\sum_{m > 0, n \leq 0} \,
 \frac{1}{E_m - E_n} \,\langle m || \vecvar{\tau} || n \rangle \nonumber \\
 &\times& \left\{ \frac{E_m + E_n}{2} \,
 \langle m || \vecvar{\tau} || n \rangle \ + \ 
 \langle m || \,\frac{1}{3} \,(\vecvar{\alpha} \cdot \vecvar{p}) \,
 \vecvar{\tau} \,|| n \rangle \right\} .
\end{eqnarray}
Since this quantity, which represents the difference of momentum 
fraction carried by the $u$-quark and the $d$-quark in the proton,
is not constrained by any conservation law, its actual value can be
estimated only numerically.

Next we turn to the discussion of the forward limit of the generalized 
magnetic form factors. First, the forward limit of 
 $G_{M, 10}^{(I = 0)} (t)$ gives
\begin{eqnarray}
 &\,& G_{M, 10}^{(I = 0)} (t = 0)
 \ = \ - \,\frac{M_N}{I} \left(\frac{N_c}{6} \right)
 \sum_{m > 0, n \leq 0} \,\frac{1}{E_m - E_n} \,
 \langle m || \vecvar{\tau} || n \rangle
 \langle m || \vecvar{r} \times \vecvar{\alpha} || n \rangle ,
\end{eqnarray}
which reproduces the known expression of the isoscalar magnetic moment of
the nucleon in the CQSM \cite{Christov95}.
On the other hand, the forward limit of 
$G_{M, 20}^{(I = 0)} (t)$ gives 
\begin{eqnarray}
 G_{M, 20}^{(I = 0)} (t = 0) 
 &=& - \,\frac{1}{I} \,\left(\frac{N_c}{6} \right) \,
 \sum_{m > 0, n \leq 0} \,\frac{1}{E_m - E_n} \,
 \langle m || \vecvar{\tau} || n \rangle \nonumber \\
 &\times& \left\{ \frac{E_m + E_n}{2} \,
 \langle m || \vecvar{r} \times \vecvar{\alpha} || n \rangle
 \ + \ \langle m || \vecvar{L} || n \rangle \right\} .
 \label{eq:GM20isFL}
\end{eqnarray}
It was shown in \cite{Ossmann05} that the rhs of the
above equation is just unity, i.e.
\begin{equation}
 G_{M,20}^{(I=0)} (t=0) \ = \ 1 . \label{eq:GM20ist0}
\end{equation}
In consideration of (\ref{eq:GM2nd}), this identity can be recast
into a little different form as
\begin{eqnarray}
 1 &=& \int_{-1}^1 x E_M^{(I = 0)} (x, 0, 0) \,d x \nonumber \\
 &=& \int_{-1}^1 x E_M^{u + d} (x, 0, 0) \,d x
 \ = \ \int_{-1}^1 x \,
 [  H^{u + d} (x, 0, 0) + E^{u + d} (x, 0, 0) ] \,d x .
\end{eqnarray}
Assuming the familiar angular momentum sum rule due to Ji
\begin{equation}
 \frac{1}{2} \int x [ H^{u + d} (x, 0, 0) + E^{u + d} (x, 0, 0) ] \,d x
 = J^{u + d} ,
\end{equation}
the above identity claims that
\begin{equation}
 J^{u + d} = \frac{1}{2} , \label{eq:TotalSpin}
\end{equation}
which means that the nucleon spin is saturated by the quark fields alone.
This is again a reasonable result, because the CQSM is an effective
quark model which contains no explicit gluon fields.
The derived identity (\ref{eq:GM20ist0}) has still another
interpretation. Remembering the fact that 
$G_{M, 20}^{(I = 0)} (t)$ consists of two parts as
\begin{equation}
 G_{M, 20}^{(I = 0)} (t) = A_{20}^{u+d} (t) + B_{20}^{u+d} (t) ,
\end{equation}
Eq.(\ref{eq:GM20ist0}) dictates that 
\begin{equation}
 A_{20}^{u + d} (0) + B_{20}^{u + d} (0) = 1 .
\end{equation}
Since it also holds that (the momentum sum rule)
\begin{equation}
 G_{M,20}^{(I = 0)} (0) = A_{20}^{u + d} (0) = 1 ,
\end{equation}
it immediately follows that
\begin{equation}
 B_{20}^{u + d} (0) = 0 ,
\end{equation}
which is interpreted as showing the absence of the {\it net quark
contribution to the anomalous gravitomagnetic moment of the nucleon}.

Finally, we investigate the forward limit of the isovector combination
of the generalized magnetic form factors. From eq. (\ref{eq:MGivform}),
we get
\begin{equation}
 G_{M, 10}^{(I = 1)} (t = 0) = 
 - \,\frac{M_N}{9} \,N_c \,\sum_{n \leq 0} \,
 \langle n || \vecvar{\tau} \cdot (\vecvar{r} \times \vecvar{\alpha} ) 
 || n \rangle ,
\end{equation}
which reproduces the known expression of the isovector magnetic form
factor of the nucleon in the CQSM. On the other hand, letting
$t \rightarrow 0$ in (\ref{eq:MG2ivform}), we have 
\begin{eqnarray}
 G_{M, 20}^{(I = 1)} (t = 0) &=& - \,\frac{1}{9} \,
 N_c \,\sum_{n \leq 0} \,
 \{ \,E_n \,\langle n || \vecvar{\tau} \cdot 
 (\vecvar{r} \times \vecvar{\alpha}) || n \rangle
 \ + \ \langle n || \vecvar{\tau} \cdot \vecvar{L} || n \rangle \} 
 \nonumber \\
 &=& \int_{-1}^1 \,x E_M^{(I = 1)} (x, 0, 0) \,d x .
\end{eqnarray}
As shown in \cite{WT05}, this sum rule can be recast into the form : 
\begin{equation}
 \frac{1}{2} \int_{-1}^1 x E_M^{(I = 1)} (x, 0, 0) \,d x 
 = J^{(I = 1)} ,
\end{equation}
where $J^{(I = 1)}$ consists of two parts as 
\begin{equation}
 J^{(I = 1)} = J_f^{(I = 1)} + \delta J^{(I = 1)} .
\end{equation}
Here, the first part is given as a proton matrix element of the
{\it free field expression} for the isovector total angular momentum
operator of quark fields as 
\begin{equation}
 J_f^{(I = 1)} = \langle p \uparrow | \hat{J}_f^{(I = 1)} | 
 p \uparrow \rangle ,
\end{equation}
with
\begin{eqnarray}
 \hat{J}_f^{(I = 1)} &=& \int \,\psi^{\dagger} (\vecvar{x}) \,
 \tau_3 \,
 \left[ (\vecvar{x} \times \hat{\vecvar{p}} )_3 + 
 \frac{1}{2} \Sigma_3 \right] \,
 \psi (\vecvar{x}) \,d^3 x \nonumber \\
 &=& \hat{L}_f^{(I = 1)} + \frac{1}{2} \,\hat{\Sigma}^{(I = 1)} .
\end{eqnarray}
On the other hand, the second term is given as
\begin{equation}
 \delta J^{(I = 1)} \ = \ - M \left(\frac{N_c}{18} \right) \,
 \sum_{n \leq 0} \,
 \langle n | r \sin F (r) \gamma^0 
 [ \vecvar{\Sigma} \cdot \hat{\vecvar{r}} \vecvar{\tau} 
 \cdot \hat{\vecvar{r}} 
 - \vecvar{\Sigma} \cdot \vecvar{\tau} ] | n \rangle .
\end{equation}

\section{Numerical Results and Discussions}

The model in the chiral limit contains two parameters, the weak pion 
decay constant $f_{\pi}$ and the dynamical quark mass $M$. As usual, 
$f_{\pi}$ is fixed to its physical value, i.e. 
$f_{\pi} = 93 \,\mbox{MeV}$.
For the mass parameter $M$, there is some argument based on the
instanton liquid picture of the QCD vacuum that it is not extremely
far from $350 \,\mbox{MeV}$ \cite{DPP88}.
The previous phenomenological analysis of various static 
baryon observables based on this model prefer a slightly larger value of 
$M$ between $350 \,\mbox{MeV}$ and $425 \,\mbox{MeV}$
\cite{Wakamatsu92R}--\cite{ARW96}.
In the present analysis, we use the value $M = 400 \,\mbox{MeV}$.
With this value of $M = 400 \,\mbox{MeV}$, we prepare self-consistent
soliton solutions for seven values of $m_{\pi}$, i.e. $m_{\pi} = 0, 100,
200, 300, 400, 500,$ and $600 \,\mbox{MeV}$, in order to see the pion
mass dependence of the generalized form factors etc.
Favorable physical predictions of the model will be obtained by using
the value of $M = 400 \,\mbox{MeV}$ and
$m_\pi = 100 \,\mbox{MeV}$, since this set gives a self-consistent
soliton solution close to the phenomenologically successful one
obtained with $M = \,375 \,\mbox{MeV}$ and $m_\pi = 0 \,\mbox{MeV}$
in the single-subtraction Pauli-Villars regularization scheme
\cite{WK98}--\cite{PPGWW99}.

\begin{figure}[htb] \centering
\begin{center}
 \includegraphics[width=9.0cm,height=7.0cm]{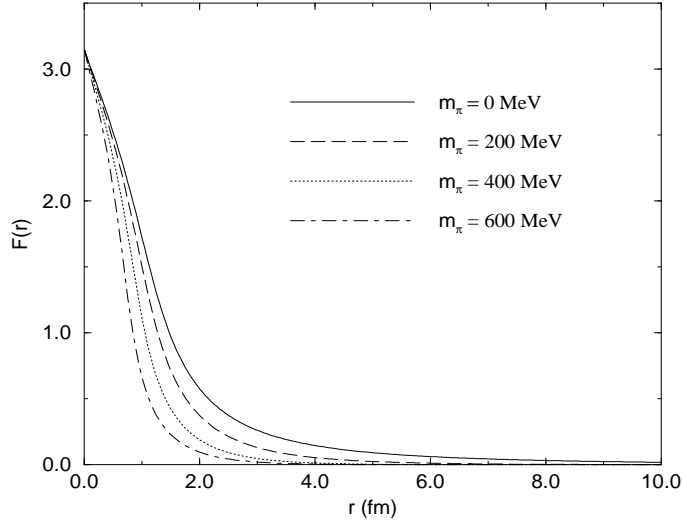}
\end{center}
\vspace*{-0.5cm}
\renewcommand{\baselinestretch}{1.20}
\caption{The self-consistent soliton profile functions obtained
with $M = 400 \,\mbox{MeV}$ and $m_\pi = 0, 200, 400$,
and $600 \,\mbox{MeV}$.}
\label{fig:profile}
\end{figure}

We first show in Fig.\ref{fig:profile}
the soliton profile functions $F (r)$ obtained with several values of 
$m_{\pi}$, i.e. $m_{\pi} = 0, 200, 400$, and $600 \,\mbox{MeV}$.
One sees that the spatial size of the soliton profile becomes more
and more compact as the pion mass increases.

We are now ready to show the theoretical predictions of the CQSM for the 
generalized form factors. Since the corresponding lattice predictions
are given for the generalized form factors $A_{n 0}^{u \pm d} (Q^2)$ and 
$B_{n 0}^{u \pm d} (Q^2)$, which are the generalization of the standard 
Dirac and Pauli form factors, we first write down the relations between 
these form factors and the generalized Sachs-type factors, which we have 
calculated in the CQSM. They are given by
\begin{eqnarray}
 A_{10}^{u + d} (t) &=& 
 \left[\, G_{E, 10}^{(I = 0)} (t) + \tau \,
 G_{M, 10}^{(I = 0)} (t) \right] \,/\, (1 + \tau) , \\
 A_{20}^{u + d} (t) &=& 
 \left[\, G_{E, 20}^{(I = 0)} (t) + \tau \,
 G_{M, 20}^{(I = 0)} (t) \right] \,/\, (1 + \tau) , \\
 A_{10}^{u - d} (t) &=& 
 \left[\, G_{E, 10}^{(I = 1)} (t) + \tau \,
 G_{M, 10}^{(I = 1)} (t) \right] \,/\, (1 + \tau) , \\
 A_{20}^{u - d} (t) &=& 
 \left[\, G_{E, 20}^{(I = 1)} (t) + \tau \,
 G_{M, 20}^{(I = 1)} (t) \right] \,/\, (1 + \tau) ,
\end{eqnarray}
and
\begin{eqnarray}
 B_{10}^{u + d} (t) &=& 
 \left[\, G_{M, 10}^{(I = 0)} (t) - G_{E, 10}^{(I = 0)} (t) \right] 
 \,/\, (1 + \tau) , \\
 B_{20}^{u + d} (t) &=& 
 \left[\, G_{M, 10}^{(I = 0)} (t) - G_{E, 20}^{(I = 0)} (t) \right] 
 \,/\, (1 + \tau) , \\
 B_{10}^{u - d} (t) &=& 
 \left[\, G_{M, 10}^{(I = 1)} (t) - G_{E, 10}^{(I = 1)} (t) \right] 
 \,/\, (1 + \tau) , \\
 B_{20}^{u - d} (t) &=& 
 \left[\, G_{M, 20}^{(I = 1)} (t) - G_{E, 20}^{(I = 1)} (t) \right] 
 \,/\, (1 + \tau) ,
\end{eqnarray}
where $\tau = - t / 4 M_N^2$.
We recall that $A_{10} (t)$ and $B_{10}(t)$ are nothing but the
standard Dirac and Pauli form factors of the nucleon :
\begin{eqnarray}
 A_{10}^{u+d} (t) &=& F_1^{(I = 0)} (t) 
 =  F_1^p (t) + F_1^n (t) , \\
 B_{10}^{u+d} (t) &=& F_2^{(I = 0)} (t) 
 =  F_2^p (t) + F_2^n (t) , \\
 A_{10}^{u-d} (t) &=& F_1^{(I = 1)} (t) 
 =  F_1^p (t) - F_1^n (t)  , \\
 B_{10}^{u-d} (t) &=& F_2^{(I = 1)} (t) 
 =  F_2^p (t) - F_2^n (t) .
\end{eqnarray}

Since the lattice simulations by the LHPC and QCDSF collaborations were 
carried out in the heavy pion region around 
$m_{\pi} \simeq (700 \sim 900) \,\mbox{MeV}$ and since the simulation
in the small pion mass region is hard to perform, we think it interesting
to investigate the pion mass dependence of the generalized form factors
within the framework of the CQSM.
For simplicity, we shall show the pion mass 
dependence of the generalized form factors at the zero momentum transfer 
only. We think it enough for our purpose because the generalized
form factors at the zero 
momentum transfer contain the most important information for clarifying
the underlying spin structure of the nucleon.
At zero momentum transfer, the 
relations between the generalized Dirac and Pauli form factors and the 
generalized Sachs-type form factors are simplified to become
\begin{eqnarray}
 A_{10}^{u + d} (0) &=& G_{E, 10}^{(I = 0)} (0) , \\ 
 A_{20}^{u + d} (0) &=& G_{E, 20}^{(I = 0)} (0) , \\ 
 A_{10}^{u - d} (0) &=& G_{E, 10}^{(I = 1)} (0) , \\ 
 A_{20}^{u - d} (0) &=& G_{E, 20}^{(I = 1)} (0) ,
\end{eqnarray}
and
\begin{eqnarray}
 B_{10}^{u + d} (0) &=& G_{M, 10}^{(I = 0)} (0)
 - G_{E, 10}^{(I = 0)} (0)  , \\ 
 B_{20}^{u + d} (0) &=& G_{M, 20}^{(I = 0)} (0)
 - G_{E, 20}^{(I = 0)} (0)  , \\ 
 B_{10}^{u - d} (0) &=& G_{M, 10}^{(I = 1)} (0)
 - G_{E, 10}^{(I = 1)} (0)  , \\ 
 B_{20}^{u - d} (0) &=& G_{M, 20}^{(I = 1)} (0)
 - G_{E, 20}^{(I = 1)} (0)  .
\end{eqnarray}

\vspace{4mm}
\begin{figure}[htb] \centering
\begin{center}
 \includegraphics[width=16.0cm,height=7.0cm]{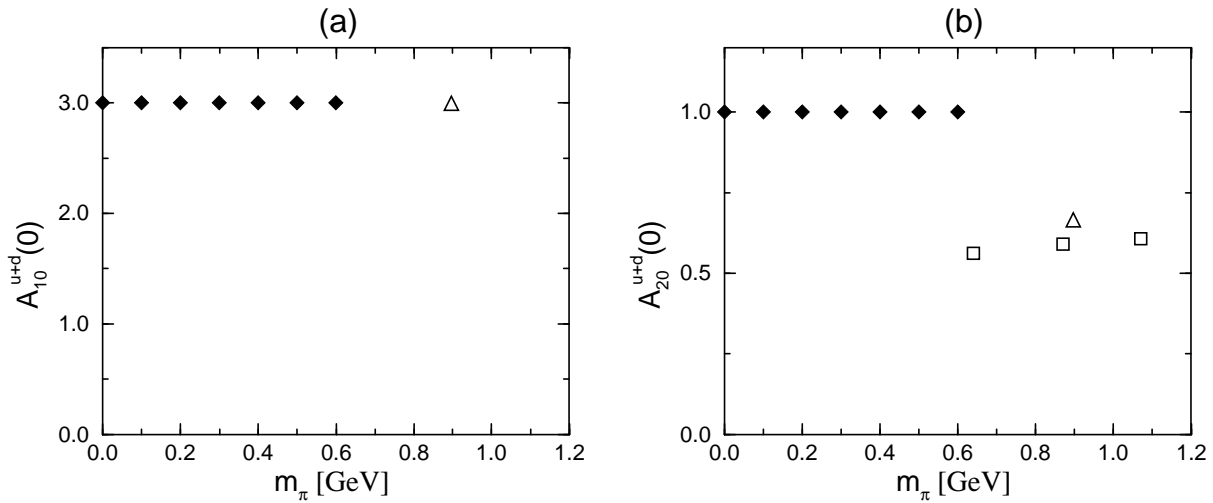}
\end{center}
\vspace*{-0.5cm}
\renewcommand{\baselinestretch}{1.20}
\caption{The predictions of the CQSM for 
$A_{10}^{u + d} (0)$ and $A_{20}^{u + d} (0)$ as functions of
$m_{\pi}$ (the filled diamonds), together with the corresponding
lattice predictions. Here, the open triangles correspond to
the predictions of the LHPC group \cite{LHPC04}, while the open
squares to those of the QCDSF collaboration \cite{QCDSF04a}.}
\label{fig:a1020is}
\end{figure}%
\vspace{2mm}

Fig.\ref{fig:a1020is} shows the predictions of the CQSM for 
$A_{10}^{u + d} (0)$ and $A_{20}^{u + d} (0)$ as functions of
$m_{\pi}$, together with the corresponding lattice predictions.
As for $A_{10}^{u + d} (0)$, the CQSM predictions and the lattice QCD
predictions are both independent of $m_{\pi}$ and consistent with
the constraint of the quark number sum rule :
\begin{equation}
 A_{10}^{u + d} (0) = N^u + N^d = 3 ,
\end{equation}
with high numerical precision. Turning to $A_{20}^{u + d} (0)$, one finds
a sizable difference between the predictions of the CQSM and of the 
lattice QCD. The lattice QCD predicts that 
\begin{equation}
 A_{20}^{u + d} (0) = \langle x \rangle^u + \langle x \rangle^d 
 \simeq (0.6 \sim 0.7) ,
\end{equation}
which means that only about $(60 \sim 70)\%$ of the total nucleon 
momentum is carried by the quark fields, while the rest is borne by the
gluon fields. On the other hand, the CQSM predictions for the same
quantity is 
\begin{equation}
 A_{20} ^{u + d} (0) = \langle x \rangle ^u + \langle x \rangle^d = 1 ,
\end{equation}
which means that the quark fields saturates the total nucleon momentum. 
This may certainly be a limitation of an effective quark model, which
contains no explicit gluon fields. Note, however, that the total quark
momentum fraction $A_{20}^{u + d} (0)$ is a scale dependent quantity.
The lattice result corresponds to the energy scale of
$Q^2 = (2 \,\mbox{GeV})^2$ \cite{LHPC04},
while the CQSM prediction should be taken as that of the model energy
scale around $Q^2 = 0.30 \,\mbox{GeV}^2 \simeq (560 \,\mbox{MeV})^2$
\cite{WK99}. We shall later make more meaningful comparison by taking
care of the scale dependencies of relevant observables.

\vspace{4mm}
\begin{figure}[htb] \centering
\begin{center}
 \includegraphics[width=16.0cm,height=7.0cm]{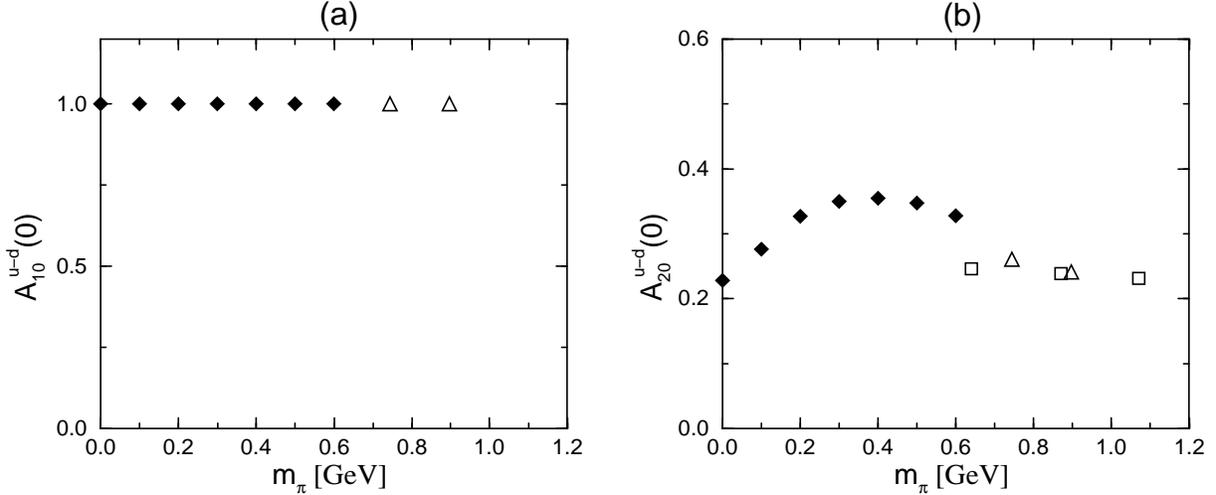}
\end{center}
\vspace*{-0.5cm}
\renewcommand{\baselinestretch}{1.20}
\caption{The predictions of the CQSM for 
$A_{10}^{u - d} (0)$ and $A_{20}^{u - d} (0)$ as functions of
$m_{\pi}$, together with the corresponding lattice
predictions \cite{LHPC04},\cite{QCDSF04a}.
The meaning of the symbols are the same as in Fig.\ref{fig:a1020is}.}
\label{fig:a1020iv}
\end{figure}%
\vspace{2mm}

Next, in Fig.\ref{fig:a1020iv}, we show the isovector combination
of the generalized form 
factors $A_{10}^{u - d} (0)$ and $A_{20}^{u - d} (0)$.
The meaning of the 
symbols are the same as in Fig.\ref{fig:a1020is}.
 As for $A_{10}^{u - d} (0)$, both the 
CQSM and the lattice simulation reproduce the quark number sum rule
\begin{equation}
 A_{10}^{u - d} (0) = N^u - N^d = 1 ,
\end{equation}
with good prediction.
Turning to $A_{20}^{u - d} (0)$, one observes that the prediction of
the CQSM shows somewhat peculiar dependence on the pion mass.
Starting from a fairly small value in the chiral limit ($m_{\pi} = 0$), 
it first increases as $m_{\pi}$ increases, but as $m_{\pi}$ further 
increases it begins to decrease, thereby showing a tendency to match the 
lattice prediction in the heavy pion region.
Very interestingly, letting put aside the absolute value,
a similar $m_{\pi}$ dependence is also observed in the chiral
extrapolation of the lattice prediction for the momentum fraction
$\langle x \rangle^u - \langle x \rangle^d$ shown in Fig.25 of 
\cite{LHPC02}.
Physically, the quantity $A_{20}^{u - d} (0)$ has a meaning of the 
difference of the momentum fractions carried by the $u$-quark and
the $d$-quark. The empirical value for it is 
$A_{20}^{u - d} (0) = \langle x \rangle ^{u - d} = 0.154 \pm 0.003$
\cite{LHPC02}. 
One sees that the prediction of the CQSM in the chiral limit is not far 
from this empirical information, although more serious comparison
must take account of the scale dependence of $\langle x \rangle^{u-d}$.

\vspace{4mm}
\begin{figure}[htb] \centering
\begin{center}
 \includegraphics[width=16.0cm,height=7.0cm]{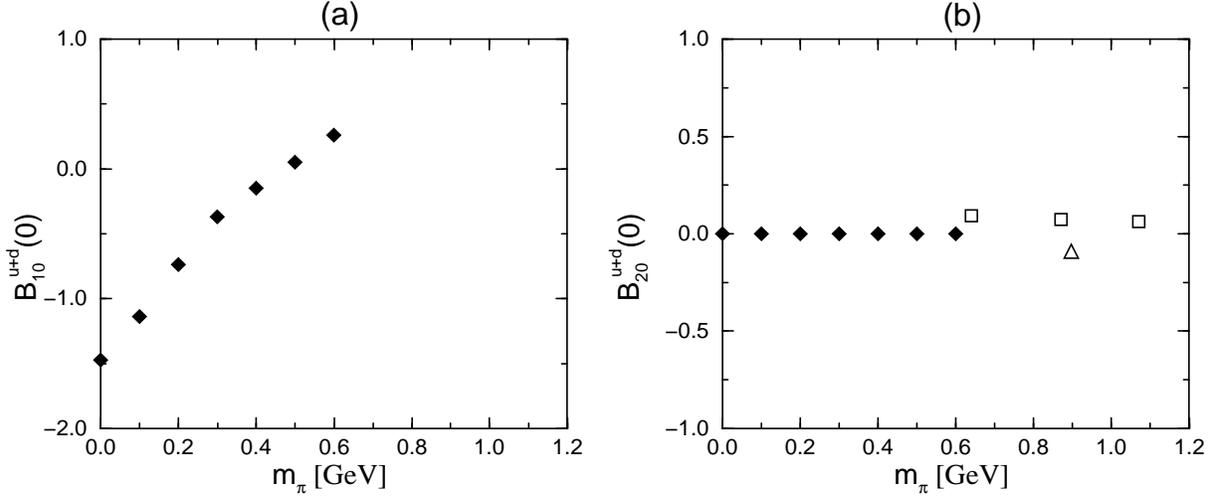}
\end{center}
\vspace*{-0.5cm}
\renewcommand{\baselinestretch}{1.20}
\caption{The predictions of the CQSM for 
$B_{10}^{u + d} (0)$ and $B_{20}^{u + d} (0)$ as functions of
$m_{\pi}$, together with the corresponding lattice
predictions \cite{LHPC05},\cite{QCDSF04b}.
The meaning of the symbols are the same as in Fig.\ref{fig:a1020is}.}
\label{fig:b1020is}
\end{figure}%

Next, shown in Fig.\ref{fig:b1020is} are the CQSM predictions for
$B_{10}^{u + d} (0)$ and $B_{20}^{u + d} (0)$.
The former quantity is related to the isoscalar 
combination of the nucleon anomalous magnetic moment as 
$B_{10}^{u + d} (0) = \kappa^u + \kappa^d = 3 \,
(\kappa^p + \kappa^n) \equiv 3 \kappa^{(T = 0)}$.
(We recall that its 
empirical value is $B_{10}^{u + d} (0) \simeq -0.36$.) We find that this 
quantity is very sensitive to the variation of the pion mass. It appears 
that the CQSM prediction $B_{10}^{u + d} (0) \simeq -1.5$ corresponding
to chiral limit underestimates the observation significantly.
However, the difference 
is exaggerated too much in this comparison. In fact, if we carry out a 
comparison in the total isoscalar magnetic moment of the nucleon 
$\frac{1}{3} \,G_{M, 10}^{(I = 0)} (0) = \mu^p + \mu^n \equiv 
\mu^{(T = 0)}$, the CQSM in the chiral limit 
gives $\mu_{CQSM}^{(T = 0)} \simeq 0.5$ in comparison with the observed 
value $\mu_{exp}^{(T = 0)} \simeq 0.88$.
To our knowledge, no theoretical 
predictions are given for this quantity by either of the LHPC or QCDSF 
collaborations. The right panel of Fig.4 shows the predictions for
$B_{20}^{u + d} (0)$, 
which is sometimes called the isoscalar part of the nucleon anomalous 
gravitomagnetic moment, or alternatively the {\it net quark contribution
to the nucleon anomalous gravitomagnetic moment}.
As already pointed out, the prediction of the CQSM for this quantity
is exactly zero :
i.e.
\begin{equation}
 B_{20}^{u + d} (0) = 0 ,
\end{equation}
The explicit numerical calculation also confirms it.
It should be recognized that the 
above result $B_{20}^{u + d} (0) = 0$ obtained in the CQSM is just
a necessary consequence of the {\it momentum sum rule} and the
{\it total nucleon spin sum rule}, both of which are saturated
by the quark field only in the CQSM as
\begin{equation}
 A_{20}^{u + d} (0) = \langle x \rangle^{u + d} = 1 ,
\end{equation}
and 
\begin{equation}
 \frac{1}{2} \,[ A_{20}^{u + d} (0) + B_{20}^{u + d} (0) ]
 \ = \ \langle J \rangle^{u + d} \ = \ \frac{1}{2} .
\end{equation}
In real QCD, the gluon also contributes to
these sum rules, thereby leading to more general identities :
\begin{eqnarray}
 &\,& A_{20}^{u + d} (0) + A_{20}^g (0) = 1 , \\
 &\,& [ A_{20}^{u + d} (0) + B_{20}^{u +d} (0) ] 
 + [ A_{20}^g (0) + B_{20}^g (0) ] = 1 ,
\end{eqnarray}
which constrains that only the sum of 
$B_{20}^{u + d} (0)$ and $B_{20}^g (0)$ is forced to vanish as
\begin{equation}
 B_{20}^{u + d} (0) + B_{20}^g (0) = 0 .
\end{equation}
(While we neglect here the contributions of other quarks than the
$u$- and $d$-quarks, it loses no generality in our discussion below. 
In fact, to include them, we have only to replace the combination 
$u + d$ by $u + d + s + \cdots$.)
The above nontrivial identity claims that the net contributions of quark
and gluon fields to the anomalous gravitomagnetic moment of the nucleon
must be zero. An interesting question is whether the quark and gluon 
contribution to the anomalous gravitomagnetic moment vanishes separately
or they are both large with opposite sign.
A perturbative analysis based on a very simple toy model indicates
the latter possibility \cite{BHMS01}.
On the other hand, a nonperturbative analysis within the framework of
the lattice QCD indicates that the net quark contribution to the
anomalous gravitomagnetic moment is small or nearly zero,
$B_{20}^{u+d}(0) = 0$ \cite{LHPC05},\cite{QCDSF04b}.
(To be more precise, we sees that the prediction of the LHPC
collaboration for $B_{20}^{u + d} (0)$ is slightly
negative \cite{LHPC05}, while that of the QCDSF group is slightly
positive \cite{QCDSF04b}.) This strongly indicates
a surprising possibility that the quark and gluon contribution to the
anomalous gravitomagnetic moment of the nucleon may separately vanish.
Worthy of special mention here is an interesting argument given by
Teryaev some years ago, claiming that the vanishing net quark
contributions to the anomalous gravitomagnetic moment of the nucleon,
violated in perturbation theory, is expected to be restored in
full nonperturbative QCD due to the confinement
\cite{Teryaev99},\cite{Teryaev98},\cite{Teryaev03}.
Very interestingly, once it
actually happens, it leads to a surprisingly
simple result, i.e. the proportionality of the quark momentum and
angular momentum fraction
\begin{equation}
 J^{u + d} = \frac{1}{2} \,\langle x \rangle^{u + d} ,
\end{equation}
as advocated by Teryaev \cite{Teryaev99},\cite{Teryaev98},\cite{Teryaev03}.
A far reaching physical consequence resulting
from this observation was extensively discussed 
in our recent report \cite{Wakamatsu05}.
(See also the discussion at the end of this section.)

\vspace{4mm}
\begin{figure}[htb] \centering
\begin{center}
 \includegraphics[width=16.0cm,height=7.0cm]{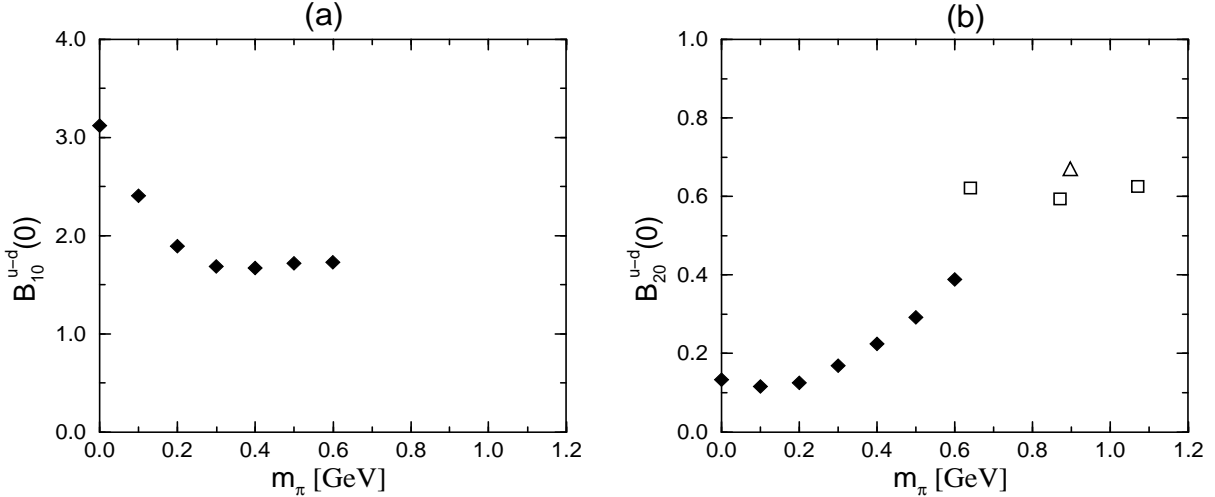}
\end{center}
\vspace*{-0.5cm}
\renewcommand{\baselinestretch}{1.20}
\caption{The predictions of the CQSM for 
$B_{10}^{u - d} (0)$ and $B_{20}^{u - d} (0)$ as functions of
$m_{\pi}$, together with the corresponding lattice
predictions \cite{LHPC05},\cite{QCDSF04b}.
The meaning of the symbols are the same as in Fig.\ref{fig:a1020is}.}
\label{fig:b1020iv}
\end{figure}%

Next, we show in Fig.\ref{fig:b1020iv} the predictions for the
isovector case, i.e. $B_{10}^{u - d} (0)$ and $B_{20}^{u - d} (0)$.
We recall first that the quantity $B_{10}^{u - d} (0)$ represents
the isovector combination of the nucleon anomalous 
magnetic moment $\kappa^{(T = 1)} \equiv \kappa^p - \kappa^n \,
(\, = \kappa^u - \kappa^d \equiv \kappa^{(I=1)} )$,
the empirical value of which is 
known to be $\kappa^{(T = 1)} = 3.706$.
One find that this quantity is extremely sensitive to the variation
of the pion mass especially near $m_\pi = 0$.
This is only natural if one remembers the important role of the pion
cloud in the isovector magnetic moment of the nucleon.
(One may notice that the prediction of the CQSM for $\kappa^{(T=1)}$
underestimates a little its empirical value even in the chiral limit.
We recall, however, that, within the framework of the CQSM, there is
an important $1 / N_c$ correction or the 1st order rotational
correction to some kind of isovector quantities
like the isovector magnetic moment of the nucleon in question or
the axial-vector coupling constant of the 
nucleon \cite{WW93}--\cite{Wakamatsu96}. This next-to-leading
correction in $1 / N_c$ should also be taken into account in more
advanced investigations.)
Shown in the right panel of Fig.\ref{fig:b1020iv} is the theoretical
predictions for $B_{20}^{u - d} (0)$,
the half of which can be interpreted as the 
difference of the total angular momentum carried by the $u$-quark and
the $d$-quark fields according to Ji's angular momentum
sum rule \cite{Ji97}.
The CQSM predicts fairly small value for this quantity,
in contrast to the lattice predictions of sizable magnitude.
It seems that the pion mass dependence 
rescues this discrepancy only partially.
Here we argue that, the reason why the CQSM (in the chiral limit)
gives rather small prediction for this quantity is intimately
connected with the characteristic $x$ dependence of the quantity
$E_M^{(I = 1)} (x, 0, 0)$, 
the forward limit of the isovector unpolarized spin-flip GPD of the
nucleon. To show it, we first recall that, within the theoretical
frame work of the CQSM, $B_{10}^{u - d} (0)$ as well as
$B_{20}^{u -d}(0)$ are calculated as 
difference of $G_{M, 10}^{(I=1)}(0)$ and $G_{E, 10}^{(I=1)}(0)$ and of 
$G_{M, 20}^{(I=1)} (0)$ and $G_{E, 20}^{(I=1)}(0)$, respectively, as 
\begin{eqnarray}
 B_{10}^{u - d} (0) &=& G_{M, 10}^{(I=1)} (0) - G_{E, 10}^{(I=1)} (0) ,\\
 B_{20}^{u - d} (0) &=& G_{M, 20}^{(I=1)} (0) - G_{E, 20}^{(I=1)} (0) .
\end{eqnarray}
Although the quantities of the rhs can be calculated directly without 
recourse to any distribution functions, they can also be evaluated as 
$x$-weighted integrals of the corresponding GPDs as 
\begin{eqnarray}
 G_{M, 10}^{(I=1)} (0) &=& \int_{-1}^1 \,
 E_M^{(I = 1)} (x, 0, 0) \,d x ,\\
 G_{M, 20}^{(I=1)} (0) &=& \int_{-1}^1 \,
 x E_M^{(I = 1)} (x, 0, 0) \,d x ,\\
 G_{E, 10}^{(I=1)} (0) &=& \int_{-1}^1 \,H_E^{(I = 1)} (x, 0, 0) \,d x
 \ = \ \int_{-1}^1 \,f^{u - d} (x) \,d x = N^u - N^d = 1 , \\
 G_{E, 20}^{(I=1)} (0) &=& \int_{-1}^1 \,x H_E^{(I = 1)} (x, 0, 0) \,d x
 \ = \ \int_{-1}^1 \,x f^{u - d} (x) \,d x 
 = \langle x \rangle^u - \langle x \rangle^d .
\end{eqnarray}
The distribution function $E_M^{(I = 1)} (x, 0, 0)$ has already been 
calculated within the CQSM in our recent paper \cite{WT05}.
As shown there, the Dirac 
sea contribution to this quantity has a sizably large peak
around $x = 0$.
Since this significant peak due to the deformed Dirac-sea quarks is
approximately symmetric with respect to the reflection
$x \longrightarrow -x$, it hardly contributes to the second moment 
$G_{M, 20}^{(I = 1)} (0)$, whereas
it gives a sizable contribution to the first moment
$G_{M, 10}^{(I = 1)} (0)$.
The predicted significant peak of $E_M^{(I = 1)} (x, 0, 0)$ around
$x = 0$ can physically be interpreted as the effects of pion cloud.
It can be convinced in several ways.
First, we investigate how this behavior of $E_M^{(I = 1)} (x, 0, 0)$ 
changes as the pion mass is varied.

\begin{figure}[htb] \centering
\begin{center}
 \includegraphics[width=9.0cm,height=7.0cm]{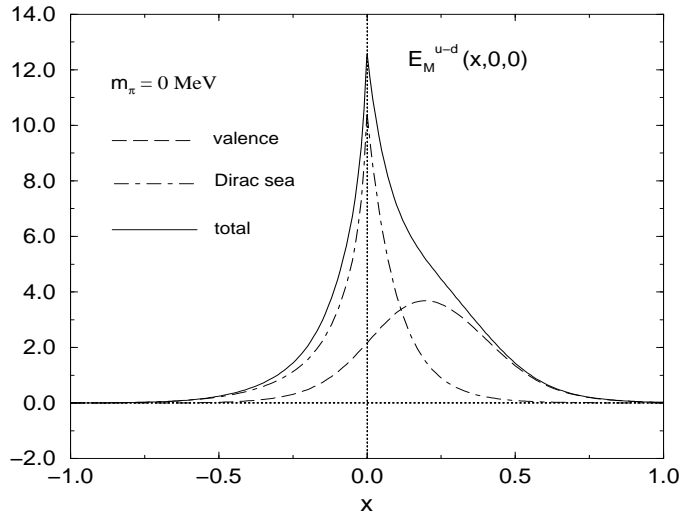}
\end{center}
\vspace*{-0.5cm}
\renewcommand{\baselinestretch}{1.20}
\caption{The prediction of the CQSM for $E_M^{u-d}(x,0,0) = 
E_M^{(I=1)}(x,0,0)$ obtained with $M = 400 \,\mbox{MeV}$ and
$m_\pi = 0$.}
\label{fig:eiv_pi0}
\end{figure}%

\vspace{6mm}
\begin{figure}[htb] \centering
\begin{center}
 \includegraphics[width=16.0cm,height=7.0cm]{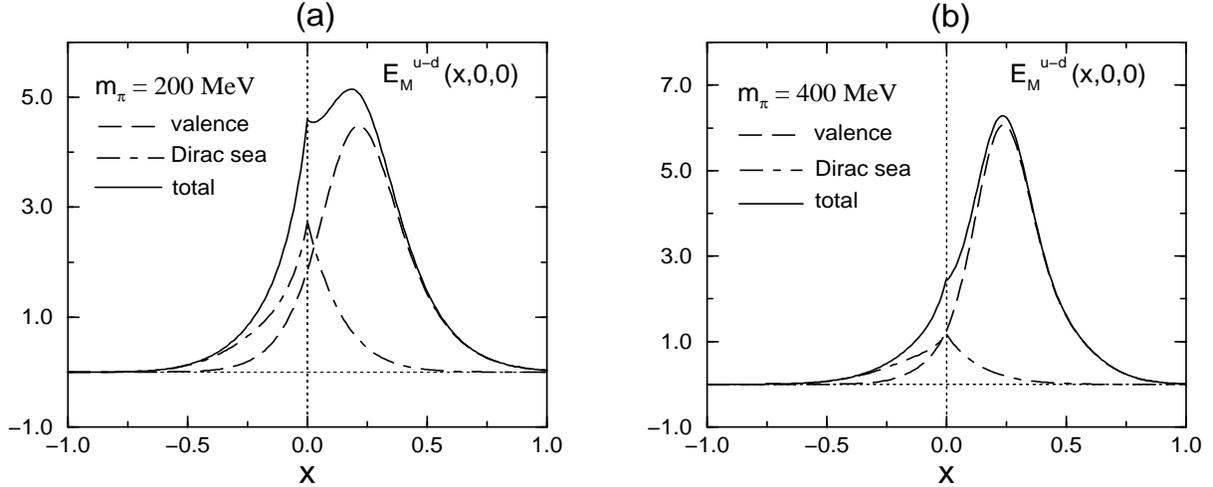}
\end{center}
\vspace*{-0.5cm}
\renewcommand{\baselinestretch}{1.20}
\caption{The $m_\pi$ dependence of $E_M^{u-d}(x,0,0)$.}
\label{fig:eiv_pi24}
\end{figure}%

Shown in Fig.\ref{fig:eiv_pi0} and in Fig.\ref{fig:eiv_pi24} are
the CQSM predictions for $E_M^{(I = 1)} (x, 0, 0)$ with several values
of $m_{\pi}$. i.e. $m_{\pi} = 0, 200$, and $400 \,\mbox{MeV}$.
One clearly sees that the height of the peak around $x = 0$, due to the 
deformed Dirac-sea quarks, decreases rapidly as $m_{\pi}$ increases.
This supports our interpretation of this peak as the effects of pion
clouds. On the other hand, one also observes that the magnitude of
the valence quark contribution, peaked around
$x \sim 1/3$, gradually increases as 
$m_{\pi}$ becomes large. This behavior of $E_M^{(I = 1)} (x, 0, 0)$
turns out to cause a somewhat unexpected $m_{\pi}$ dependence of 
$G_{M, 10}^{(I = 1)} (0)$ and $G_{M, 20}^{(I = 1)} (0)$.
As a function of $m_{\pi}$, the Dirac sea contribution to
$G_{M, 10}^{(I = 1)} (0)$ decreases fast,
whereas the valence quark contribution to it increases slowly, so 
that the total $G_{M, 10}^{(I = 1)} (0)$ becomes a decreasing function
of $m_{\pi}$.
On the other hand, owing to the approximate odd-function nature of
the Dirac sea contribution to $x E_M^{(I = 1)} (x, 0, 0)$ with respect
to $x$, it hardly contributes to $G_{M, 20}^{(I = 0)} (0)$ independent
of the pion mass, while the valence quark contribution to 
$x E_M^{(I = 1)} (x, 0, 0)$ is an increasing function of $m_{\pi}$, 
thereby leading to the result that the net $G_{M, 20}^{(I = 0)} (0)$ is 
a increasing function of $m_{\pi}$. 

\begin{figure}[htb] \centering
\begin{center}
 \includegraphics[width=9.0cm,height=7.0cm]{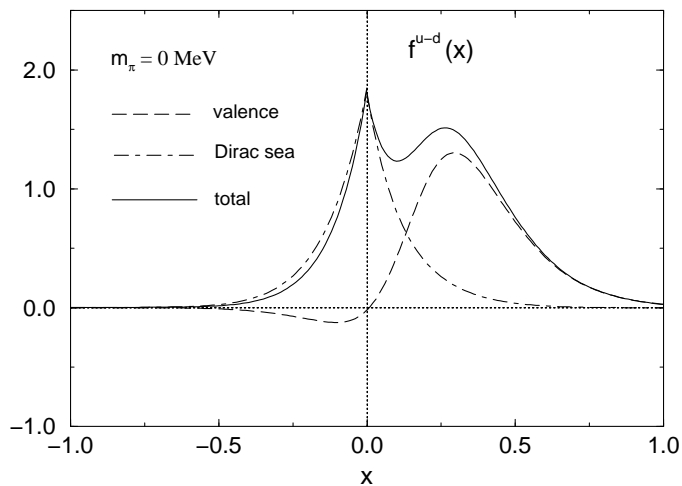}
\end{center}
\vspace*{-0.5cm}
\renewcommand{\baselinestretch}{1.20}
\caption{The prediction of the CQSM for $f^{u-d}(x,0,0)$
obtained with $M = 400 \,\mbox{MeV}$ and $m_\pi = 0$.}
\label{fig:fiv_pi0}
\end{figure}%

\vspace{6mm}
\begin{figure}[htb] \centering
\begin{center}
 \includegraphics[width=16.0cm,height=7.0cm]{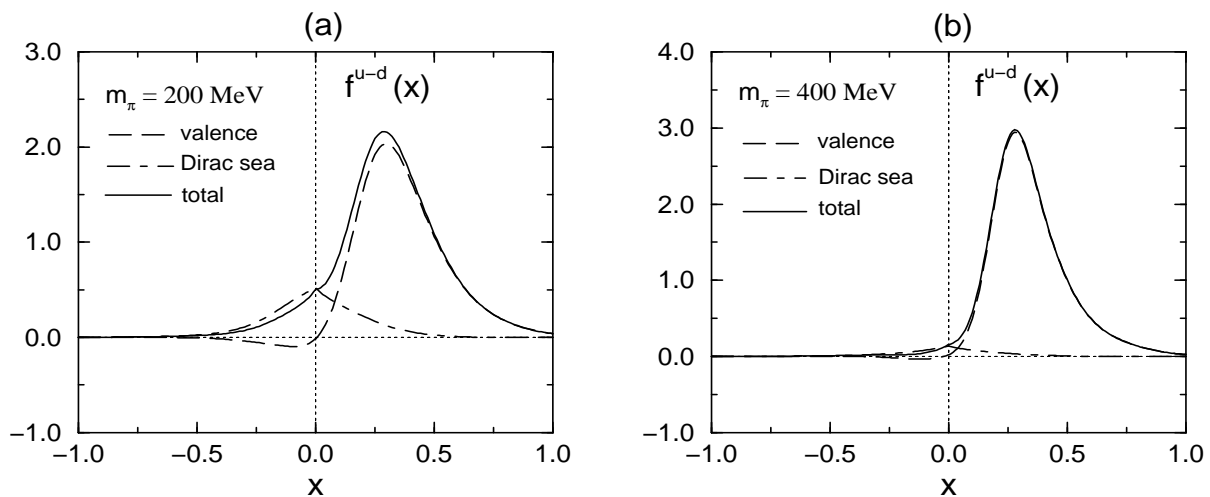}
\end{center}
\vspace*{-0.5cm}
\renewcommand{\baselinestretch}{1.20}
\caption{The $m_\pi$ dependence of $f^{u-d}(x,0,0)$.}
\label{fig:fiv_pi24}
\end{figure}%
\vspace{2mm}

We can give still another support to the above-mentioned interpretation
of the contribution of the Dirac-sea quarks. To see it, we first recall 
that the theoretical unpolarized distribution function $f^{u - d} (x)$
appearing in the decomposition
\begin{equation}
 E_M^{u-d} (x,0,0) = f^{u-d}(x) + E^{u-d}(x,0,0), 
\end{equation}
also has a sizable peak around $x = 0$ due to the deformed Dirac-sea
quarks.
As shown in Fig.\ref{fig:fiv_pi0} and in Fig.\ref{fig:fiv_pi24},
this peak is again a rapidly decreasing function 
of $m_{\pi}$, supporting our interpretation of it as the effects of pion 
clouds. Here, we can say more. We point out that this small-$x$ behavior of $f^{u - d} (x)$ is just what is required by the famous
NMC measurement \cite{NMC91}.
To confirm it, first remember that the distribution $f^{u - d}(x)$ in the
negative $x$ region should actually be interpreted as the distribution of
antiquarks. To be explicit, it holds that
\begin{equation}
 \bar{u} (x) - \bar{d} (x) = -f^{u - d} (-x) \ \ \ (x \geq 0) .
\end{equation}
The large and positive value of $f^{u - d} (x)$ in the negative $x$
region close to $x = 0$ means that $\bar{u}(x) - \bar{d} (x)$ is
negative, i.e. the dominance of the $\bar{d}$-quark over
the $\bar{u}$-quark inside the proton, which has been established
by the NMC measurement \cite{NMC91}.

\vspace{4mm}
\begin{figure}[htb] \centering
\begin{center}
 \includegraphics[width=9.0cm,height=7.5cm]{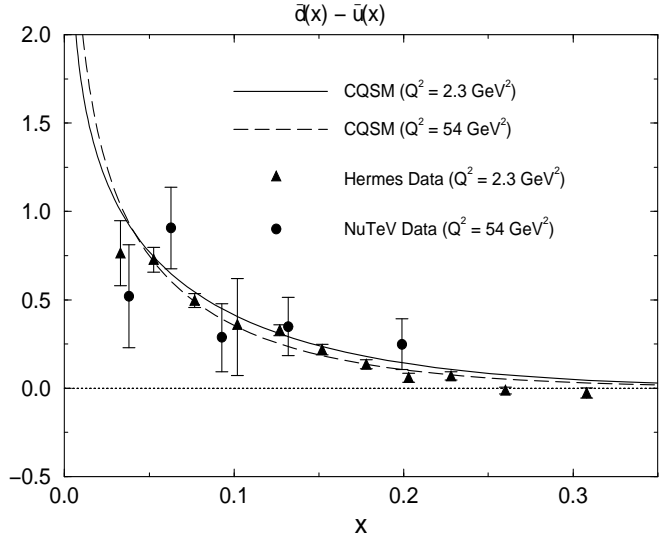}
\end{center}
\vspace*{-0.5cm}
\renewcommand{\baselinestretch}{1.20}
\caption{The prediction of the CQSM for $\bar{d}(x) - \bar{u}(x)$
evolved to $Q^2 = 2.3 \,\mbox{GeV}^2$ and $Q^2 = 54 \,\mbox{GeV}^2$
in comparison with the Hermes and NuTeV data at the corresponding
energy scales \cite{HERMES98},\cite{E866}.}
\label{fig:dbdu}
\end{figure}%
\vspace{3mm}

Shown in Fig.\ref{fig:dbdu} are the predictions of the CQSM
for $\bar{d} (x) - \bar{u} (x)$ evolved to the high energy scales
corresponding to the experimental observation \cite{Saga96}.
(The theoretical predictions here were obtained with
$M = 400 \,\mbox{MeV}$ and $m_{\pi} = 100 \,\mbox{MeV}$.)
The model reproduces well the observed behavior of 
$\bar{d} (x) - \bar{u} (x)$, although the magnitude of the flavor
asymmetry in smaller $x$ region seems to be slightly overestimated.
It is a widely accepted fact that this flavor 
asymmetry of the sea quark distribution in the proton can physically be 
understood as the effects of pion cloud at least qualitatively
\cite{HM91}--\cite{Wakamatsu92}.
This then supports our interpretation of the effects of the deformed
Dirac-sea quarks in $E_M^{(I = 1)} (x, 0, 0)$ and $f^{u - d} (x)$
as the effects of pion clouds.

\begin{table}[htbp]
\begin{center}
\renewcommand{\baselinestretch}{1.20}
\caption{The $m_\pi$ dependencies of $G_{M,20}^{(I=1)}(0)$,
$G_{E,20}^{(I=1)}(0)$, and $B_{20}^{u-d} (0)$ in the CQSM with
$M = 400 \,\mbox{MeV}$. \label{tabivb20}}
\vspace{10mm}
\begin{tabular}{cccc} \hline
 \ \ $m_\pi \, (\mbox{MeV})$ \ \ & \ \ \ $G_{M,20}^{(I=1)}(0)$ \ \ \ & 
 \ \ \ $G_{E,20}^{(I=1)}(0)$ \ \ \ & \ \ \ $B_{20}^{u-d} (0)$ \ \ \ 
 \\ \hline\hline
 \   0 \ & 0.361 & 0.228 & 0.133 \\ \hline
 \ 100 \ & 0.392 & 0.276 & 0.116 \\ \hline
 \ 200 \ & 0.452 & 0.327 & 0.125 \\ \hline
 \ 300 \ & 0.519 & 0.350 & 0.169 \\ \hline
 \ 400 \ & 0.579 & 0.354 & 0.225 \\ \hline
 \ 500 \ & 0.640 & 0.347 & 0.293 \\ \hline
 \ 600 \ & 0.716 & 0.328 & 0.388 \\ \hline
\end{tabular}
\end{center}
\renewcommand{\baselinestretch}{1.20}
\end{table}

We show in Table \ref{tabivb20} the model predictions for 
$G_{M, 20}^{(I = 1)} (0), G_{E, 20}^{(I = 1)} (0)$, and 
$B_{20}^{u - d} (0) = G_{M, 20}^{(I = 1)} (0) - G_{E, 20}^{(I = 1)} (0)$ 
as functions of $m_{\pi}$.
One sees that the value of $G_{M, 20}^{(I = 1)} (0)$ with $m_{\pi} = 0$
is an order of $0.3 \sim 0.4$.
As already pointed out, it gradually increases as $m_{\pi}$ becomes
large. This is also the case with $G_{E, 20}^{(I = 1)} (0)$.
As a consequence, the isovector combination of the nucleon anomalous 
gravitomagnetic moment $B_{20}^{u - d} (0)$, which is obtained as a 
difference of the above two quantities, is also an increasing function
of $m_{\pi}$, thereby having a tendency to come closer to the lattice
prediction given in the heavy pion region. Still, the CQSM predictions 
$B_{20}^{u - d} (0) \simeq 0.3$ around $m_{\pi} = 500 \,\mbox{MeV}$ is a factor of two smaller than the corresponding lattice prediction 
$B_{20}^{u - d} (0) \simeq 0.6$.
Now we summarize the reason why the CQSM gives fairly 
small prediction for $B_{20}^{u -d} (0)$. It is due to two types of 
cancellations. The first is the cancellation of the potentially large 
contribution of Dirac-sea quarks arising from the approximately 
antisymmetric behavior of the Dirac sea contribution to 
$x E_M^{(I = 1)} (x, 0, 0)$ as well as $x f^{u - d} (x)$.
The second is the cancellation between the total gravitomagnetic moment 
$G_{M, 20}^{(I = 1)} (0)$ and its canonical part
$G_{E, 20}^{(I = 1)} (0)$.
We are not sure whether the lattice simulation carried out in the heavy 
pion region with neglect of the so-called disconnected diagrams can 
efficiently take account of such effects of chiral dynamics as discussed 
above.

\begin{figure}[htb] \centering
\begin{center}
 \includegraphics[width=16.0cm,height=7.5cm]{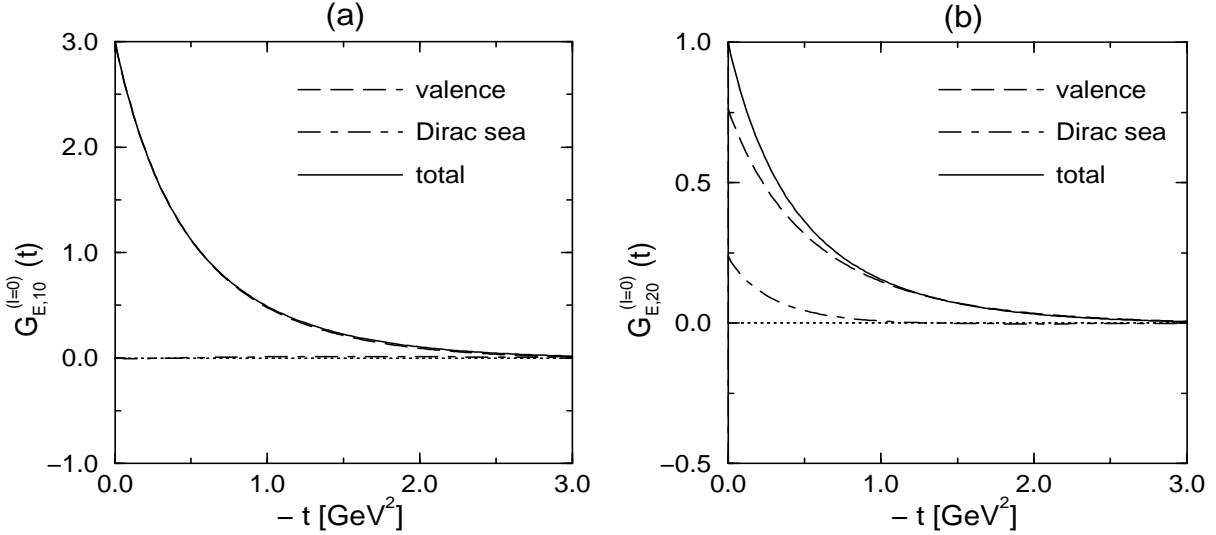}
\end{center}
\vspace*{-0.5cm}
\renewcommand{\baselinestretch}{1.20}
\caption{The predictions of the CQSM for the isoscalar generalized electric
form factors $G^{(I=0)}_{E,10}(t)$ and $G^{(I=0)}_{E,20}(t)$.}
\label{fig:GEis1020}
\end{figure}

\begin{figure}[htb] \centering
\begin{center}
 \includegraphics[width=16.0cm,height=7.5cm]{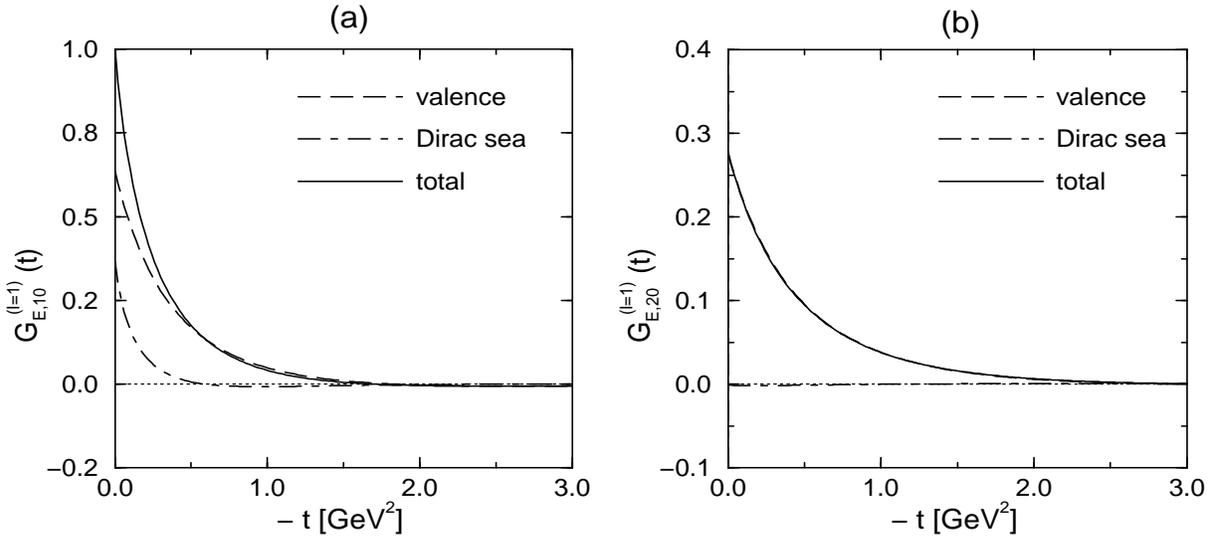}
\end{center}
\vspace*{-0.5cm}
\renewcommand{\baselinestretch}{1.20}
\caption{The predictions of the CQSM for the isovector generalized electric
form factors $G^{(I=1)}_{E,10}(t)$ and $G^{(I=1)}_{E,20}(t)$.}
\label{fig:GEiv1020}
\end{figure}

\begin{figure}[htb] \centering
\begin{center}
 \includegraphics[width=16.0cm,height=7.5cm]{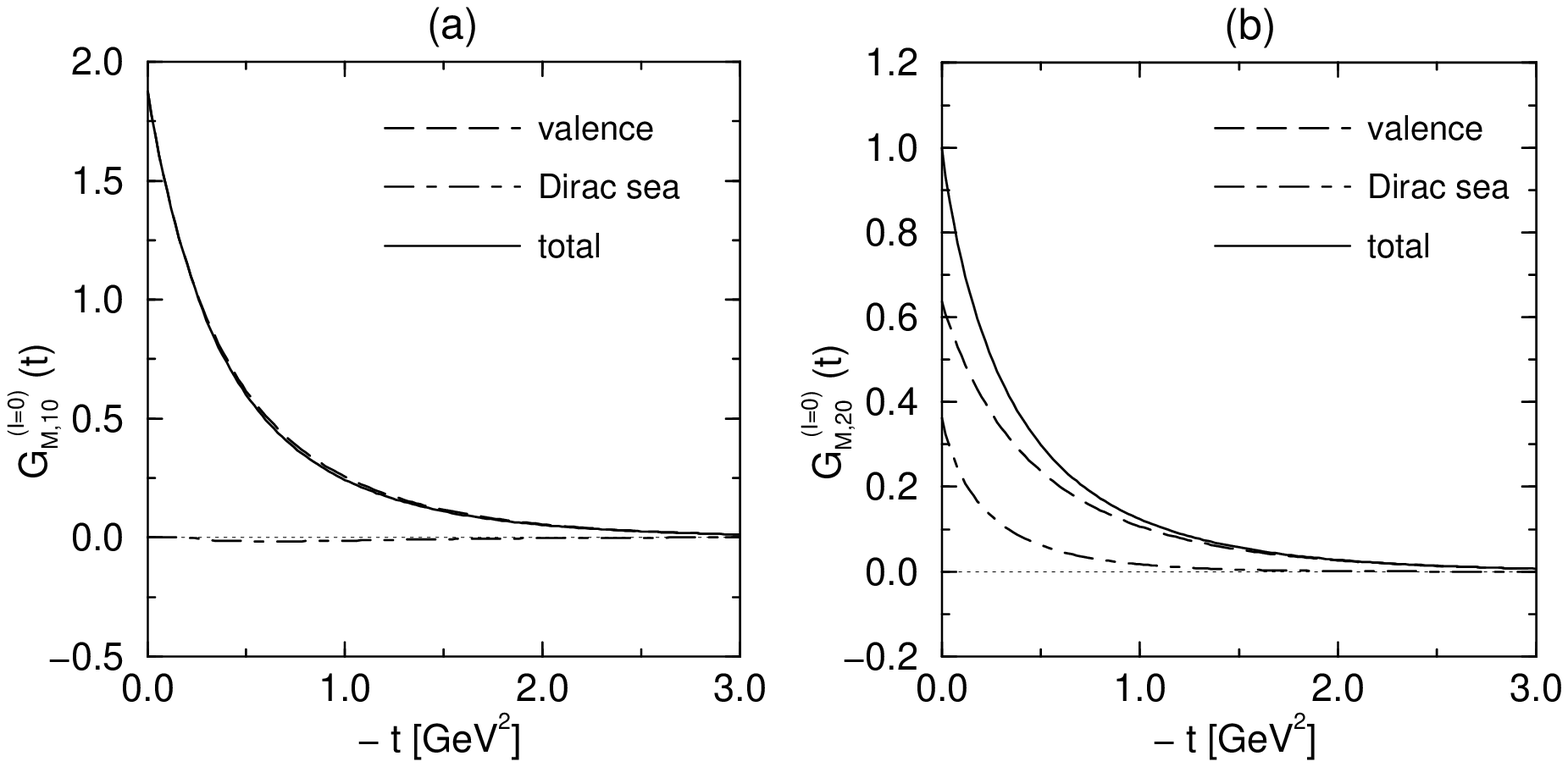}
\end{center}
\vspace*{-0.5cm}
\renewcommand{\baselinestretch}{1.20}
\caption{The predictions of the CQSM for the isoscalar generalized
magnetic form factors $G^{(I=0)}_{M,10}(t)$ and $G^{(I=0)}_{M,20}(t)$.}
\label{fig:GMis1020}
\end{figure}

\begin{figure}[htb] \centering
\begin{center}
 \includegraphics[width=16.0cm,height=7.5cm]{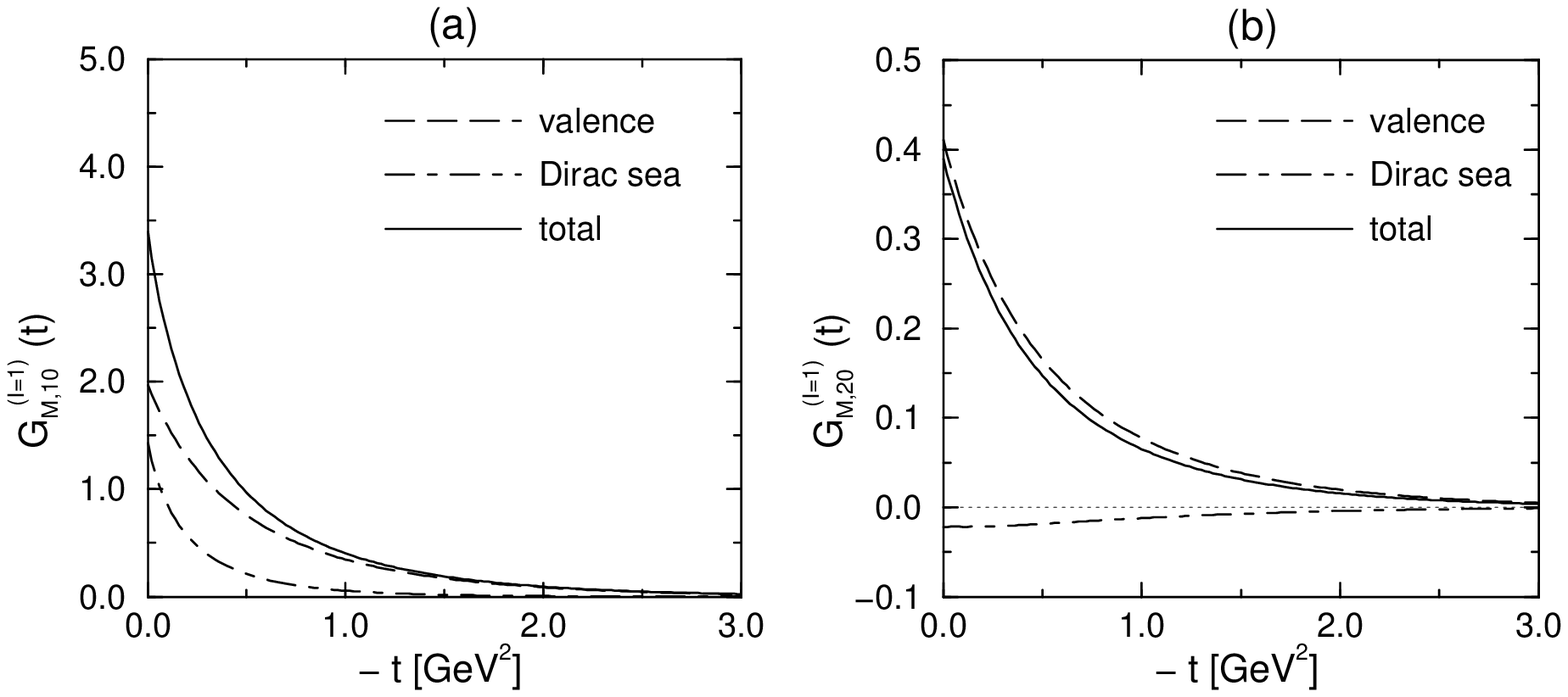}
\end{center}
\vspace*{-0.5cm}
\renewcommand{\baselinestretch}{1.20}
\caption{The predictions of the CQSM for the isovector generalized magnetic
form factors $G^{(I=1)}_{M,10}(t)$ and $G^{(I=1)}_{M,20}(t)$.}
\label{fig:GMiv1020}
\end{figure}

So far, we have investigated the pion mass dependence of the
forward limit of the generalized form factors of the nucleon.
Here, we investigate the momentum-transfer dependencies of some
form factors of the nucleon.
For the reason explained before,
all the physical predictions given hereafter will be 
obtained with use of the mass parameter $M = 400 \,\mbox{MeV}$
and $m_{\pi} = 100 \,\mbox{MeV}$. 
We show in Fig.\ref{fig:GEis1020}, Fig.\ref{fig:GEiv1020},
Fig.\ref{fig:GMis1020}, Fig.\ref{fig:GMiv1020} the predicted
momentum-transfer dependencies of the generalized Sachs form factors,
$G_{E, 10}^{(I = 0)} (t)$, $G_{E, 20}^{(I = 0)} (t)$,
$G_{E, 10}^{(I = 1)} (t)$, $G_{E, 20}^{(I = 1)} (t)$,
$G_{M, 10}^{(I = 0)} (t)$, $G_{M, 20}^{(I = 0)} (t)$,
$G_{M, 10}^{(I = 1)} (t)$, and $G_{M, 20}^{(I = 1)} (t)$.


\vspace{4mm}


\vspace{4mm}


\begin{figure}[htb] \centering
\begin{center}
 \includegraphics[width=16.0cm,height=7.5cm]{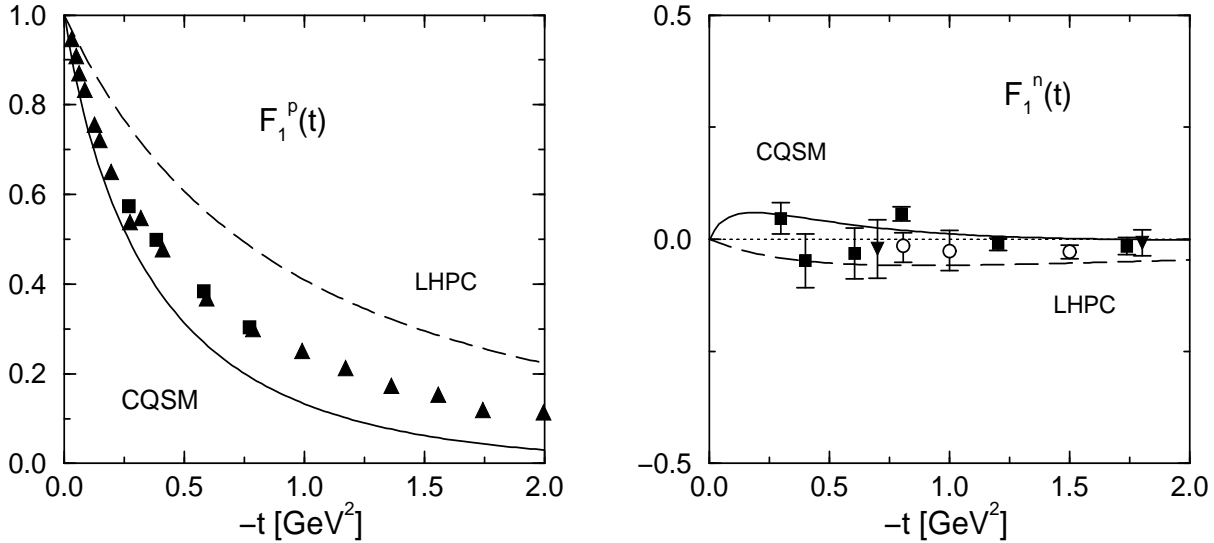}
\end{center}
\vspace*{-0.5cm}
\renewcommand{\baselinestretch}{1.20}
\caption{The prediction of the CQSM for the Dirac form factors
of the proton and the neutron in comparison with the empirical data
\cite{Arnold86} together with the LHPC lattice predictions.
The LHPC lattice predictions are based on their dipole fits
\cite{LHPC04},\cite{LHPC05}.}
\label{fig:F1pn}
\end{figure}

To get some feeling about the momentum-transfer dependencies of the 
predicted form factors, we shall compare them with the existing empirical
data.
At present, only the lowest moment of the GPDs, i.e. the standard 
electromagnetic form factors of the nucleon are experimentally known.
Shown in Fig.\ref{fig:F1pn} are the predictions of the CQSM for the
Dirac form factors of the proton and the neutron,
in comparison with the empirical data \cite{Arnold86}
together with the corresponding predictions at the LHPC lattice
simulation \cite{LHPC04},\cite{LHPC05}.
One observes that the $t$-dependence of the CQSM prediction
for $F_1^p (t)$ is a little too stronger than the empirical one,
while the t-dependence of the lattice predictions is too much weaker
than the empirical one.
A little too fast falloff of the CQSM predictions means that it 
slightly overestimates the electromagnetic size of the proton.
On the contrary, the electromagnetic proton size predicted by the LHPC 
simulation is too small as compared with the empirically known size.
As is well known, the Dirac form factor of the neutron is not well 
determined experimentally. Both of the CQSM prediction and the lattice 
QCD predictions are very small in magnitude in qualitatively consistent 
with the empirical information, although the former is slightly
positive, while the latter is slightly negative.

\vspace{4mm}

\begin{figure}[htb] \centering
\begin{center}
 \includegraphics[width=10.0cm,height=7.5cm]{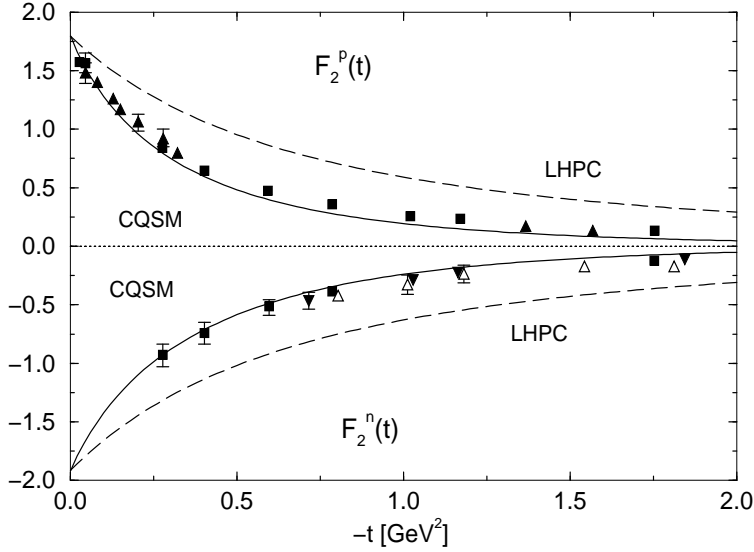}
\end{center}
\vspace*{-0.5cm}
\renewcommand{\baselinestretch}{1.20}
\caption{The predictions of the CQSM for the Pauli form factors
of the proton and the neutron in comparison with the empirical data
\cite{Arnold86} together with the LHPC lattice predictions based on
the dipole fits \cite{LHPC04},\cite{LHPC05}.}
\label{fig:F2pn}
\end{figure}

Fig.\ref{fig:F2pn} shows the predictions of the 
CQSM for the Pauli form factors of the proton and the neutron, in 
comparison with the empirical data together with the corresponding 
predictions of the LHPC lattice simulation. (Here, both of the CQSM
predictions and the lattice QCD predictions are normalized to the
observed anomalous magnetic moments of the proton and the neutron
at the zero momentum transfer.)
The solid curves represent the predictions of the CQSM, whereas the
dashed curves show the corresponding lattice predictions.
One can see that the 
predictions of the CQSM reproduce the empirical Pauli form factors
fairly well.
On the other hand, the lattice QCD predicts too slow falloff of the 
Pauli form factors, which means that the magnetic sizes of the nucleon are
largely underestimated by the lattice QCD simulations. The underestimate 
of the nucleon electromagnetic sizes seems to be a general tendency of
the lattice QCD simulations by the LHPC and QCDSF collaborations.
It is not clear yet whether the origin of discrepancy can be traced
back to the fact that these lattice simulations were carried out in
the region with unrealistically heavy pion mass.

\vspace{4mm}

\begin{figure}[htb] \centering
\begin{center}
 \includegraphics[width=10.0cm,height=7.5cm]{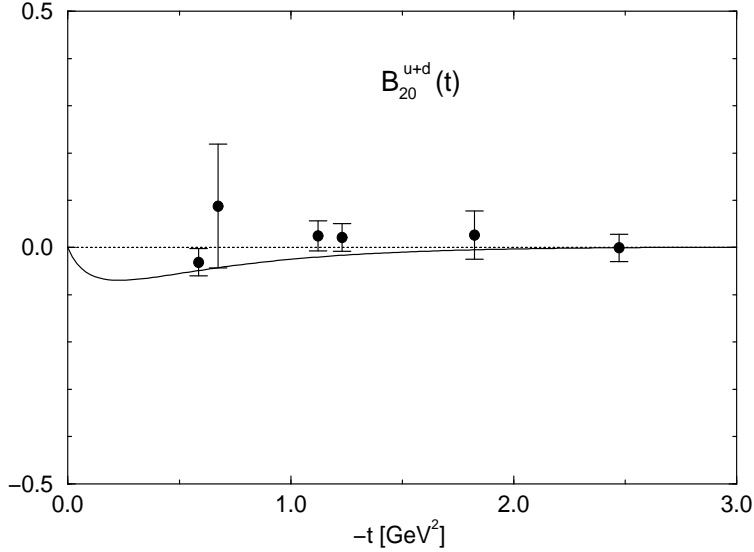}
\end{center}
\vspace*{-0.5cm}
\renewcommand{\baselinestretch}{1.20}
\caption{The predictions of the CQSM for the generalized form factor
$B_{20}^{u+d} (t)$ of the nucleon, in comparison with the
corresponding prediction of the lattice QCD by LHPC
collaboration \cite{LHPC05}.}
\label{fig:B20t}
\end{figure}

Concerning the genuine generalized form factors with $n \geq 2$,
we have no experimental information yet. Among them, of our
particular interest is the generalized form factor $B_{20}^{u+d}(t)$
appearing in Ji's angular momentum sum rule. We show in 
Fig.\ref{fig:B20t} the prediction of the CQSM for $B_{20}^{u+d}(t)$
obtained with $M = 400 \,\mbox{MeV}$ and $m_\pi = 100 \,\mbox{MeV}$,
in comparison with the corresponding predictions of the LHPC group.
One sees that, for arbitrary value of $t$, $B_{20}^{u+d}(t)$ does not
vanish in the CQSM, which is an indication of the fact that the shapes
of the quark momentum and the total angular momentum distributions
are not completely the same. Still, the magnitude of $B_{20}^{u+d}(t)$
turns out to be very small, which seems qualitatively compatible
with the prediction of the LHPC group, although one should not
forget about large uncertainties in the lattice simulation at the
present level.

\vspace{4mm}

We are now ready to discuss what we can say about the spin contents of
the nucleon from our investigation on the generalized form factors of
the nucleon.
The quantities of our interest are all obtained from the forward limit of
the generalized form factors, which are defined as the second moments
of the relevant GPDs : 
\begin{eqnarray}
 \langle x \rangle^{u + d} &=& G_{E, 20}^{(I = 0)} (0), \ \ \ \ \ 
 \langle x \rangle^{u - d} \ = \ G_{E, 20}^{(I = 1)} (0), \\
 2 \,J^{u + d} &=& G_{M, 20}^{(I = 0)} (0), \ \ \ \ \ 
 2 \,J^{u - d} \ = \ G_{M, 20}^{(I = 1)} (0) .
\end{eqnarray}
Summarized below are the predictions of the CQSM model for these quantities obtained with $M = 400 \,\mbox{MeV}$ and $m_{\pi} = 100 \,\mbox{MeV}$ :
\begin{eqnarray}
 \langle x \rangle^{u + d} &=& 1.00, \ \ \ \ \ 
 \langle x \rangle^{u - d} = 0.276, \\
 2 J^{u + d} &=& 1.00, \ \ \ \ \ 
 2 J^{u - d} = 0.406 .
\end{eqnarray}
As pointed out before, these second moments of GPDs are generally scale 
dependent. Our viewpoint is that the predictions of the CQSM correspond
to those at the low energy scale where the validity of the model is
ensured. (This energy may typically be characterized by the Pauli-Villars
mass $\Lambda_1 \simeq 600 \,\mbox{MeV}$.) 
We shall take account of the scale 
dependence of the above quantities by solving the QCD evolution
equation at the next-to-leading order (NLO) with the predictions of
the CQSM as the initial conditions \cite{FRS77}--\cite{ABY97}. 
For simplicity, let us assume that, at this low energy 
scale, there is no contribution of gluon fields as well as of the
strange quarks, which dictates that
\begin{equation}
 \langle x \rangle^s = 0.0, \ \ \ 2 \,J^s = 0.0, \ \ \ \ 
 \langle x \rangle^g = 0.0, \ \ \ 2 \,J^g = 0.0 .
\end{equation}
The starting energy of the evolution is taken to be 
$Q_{ini}^2 = 0.30 \,\mbox{GeV}^2 \simeq (550 \,\mbox{MeV})^2$,
because it is favored from the previous successful application of
the model to high energy deep-inelastic-scattering
observables \cite{WK98}--\cite{Wakamatsu03b}.
Taking $N_f = 3$ and $\Lambda_{QCD} = 0.248 \,\mbox{GeV}$,
we find that, at $Q^2 = 4 \,\mbox{GeV}^2$, 
\begin{eqnarray}
 \langle x \rangle^{u + d + s} &=& 0.676, \ \ \ 
 \langle x \rangle^{u - d} = 0.171, \ \ \ 
 \langle x \rangle^g = 0.324, \\
 2 J^{u + d + s} &=& 0.676, \ \ \ 
 2 J^{u - d} = 0.257, \ \ \ \,
 2 J^g = 0.324 .
\end{eqnarray}
One may notice that the values of $\langle x \rangle^{u+d+s}$ and
$2 \,J^{u+d+s}$ at $Q^2 = 4 \,\mbox{GeV}^2$ precisely coincide.
Actually, the equality of these two quantities holds at any energy
scale. The reason is because these two quantities obey exactly the
same evolution equation and because they are equal at the initial
energy scale according to the CQSM \cite{JTH96}.
(We emphasize that the latter is
also the case for the LHPC and QCDSF lattice predictions at least
approximately \cite{LHPC05},\cite{QCDSF04b}.)
In table \ref{tabxandj}, we compare these predictions of the CQSM
with those of the lattice QCD and also with the empirical values for 
$\langle x \rangle^{u + d + s}$ and $\langle x \rangle^{u - d}$ obtained
from the phenomenological PDF fits \cite{MRST2004}.
As one sees, the total momentum
fraction $\langle x \rangle^{u + d + s}$
carried by the quarks decreases rapidly as 
$Q^2$ increases. The evolved value $\langle x \rangle ^{u + d} 
\simeq 0.68$ at $Q^2 = 4 \,\mbox{GeV}^2$ is not extremely far from the
lattice QCD prediction at the same normalization scale,
although it is a little larger than the empirical value
$\langle x \rangle^{u + d + s}_{empirical} \simeq 0.57$. This difference 
may be an indication of the fact that, even at the low energy scale
around $Q^2 \simeq 0.30 \,\mbox{GeV}^2$,
the gluons may carry some portion of the
nucleon momentum. As far as the difference of the momentum fractions
carried by the $u$-quark and the $d$-quark, the CQSM well reproduces
the empirical value $\langle x \rangle^{u - d}_{empirical} \simeq 0.16$
at $Q^2 = 4 \, \mbox{GeV}^2$, whereas the lattice QCD overestimates it
a little.

\begin{table}[htbp]
\begin{center}
\renewcommand{\baselinestretch}{1.20}
\caption{The predictions of the CQSM for ${\langle x \rangle}^{u+d}$,
${\langle x \rangle}^{u-d}$, $2 \,J^{u+d}$, and $2 \,J^{u-d}$
in comparison with the predictions of the lattice QCD
simulations \cite{LHPC04},\cite{LHPC05},\cite{QCDSF04b},\cite{QCDSF05}
as well as with the empirical 
information \cite{MRST2004}. \label{tabxandj}}
\vspace{10mm}
\begin{tabular}{cccccc} \hline
 \, & \ \ CQSM (model scale) \ \ & \ \ 
 CQSM \,($Q^2 = 4 \,\mbox{GeV}^2)$ \ \ & 
 \ \ LHPC \ \ & \ \ QCDSF \ \ & \ \ empirical \ \ 
 \\ \hline\hline
 \ ${\langle x \rangle}^{u+d+s}$ \ & 1.000 & 0.676 & 0.61 & 0.59 & 0.57
 \\ \hline
 \ ${\langle x \rangle}^{u-d}$ \ & 0.276 & 0.171 & 0.269 & 0.24 & 0.157
 \\ \hline
 \ $2 \,J^{u+d+s}$ \ & 1.000 & 0.676 & 0.58 & 0.66 & --- 
 \\ \hline
 \ $2 \,J^{u-d}$ \ & 0.406 & 0.257 & 0.93 & 0.82 & --- 
 \\ \hline
\end{tabular}
\end{center}
\renewcommand{\baselinestretch}{1.20}
\end{table}

Next we turn to the discussion of the total angular momenta
$J^u$ and $J^d$ carried by the quark fields, on which we do not have
any empirical information yet.
One can see that, as far as the total angular momentum fraction 
$J^{u + d + s}$ carried by the quark fields, is concerned, the prediction
of the CQSM is qualitatively consistent with that of the lattice QCD.
However, a big discrepancy is observed for the difference $J^{u - d}$
of the angular momentum carried by the $u$-quark and the $d$-quark.
The cause of this discrepancy can be traced back to that of the isovector
gravitomagnetic moment of the nucleon $G_{M, 20}^{(I = 1)} (0)$,
which we have already discussed.
Fairly small prediction of the CQSM for $J^{u - d}$ also appears to be 
incompatible with the semi-theoretical (or semi-phenomenological)
estimate carried out in \cite{GPV01} with partial use of the predictions
of the CQSM.
As we shall discuss below, however, their estimate for
$J^u$ and $J^d$ shown in Table 4 and Table 5 of \cite{GPV01}
should be taken with care. In fact, it was
obtained based on the valence-like approximation, i.e. by neglecting the
sizable Dirac sea contributions to $E^u (x, 0, 0)$ and $E^d (x, 0, 0)$.
To be more concrete, they start with a simple guess for these
distributions functions as 
\begin{eqnarray}
 E^u (x, 0, 0) &=& \frac{1}{2} \,\kappa^u \,f^{u_{val}} (x), \\
 E^d (x, 0, 0) &=& \ \ \kappa^d \,f^{d_{val}} (x) ,
\end{eqnarray}
with 
\begin{eqnarray}
 \kappa^u &=& 2 \,\kappa^p + \kappa^u = \ 1.673, \\
 \kappa^d &=& \kappa^p + 2 \,\kappa^u = - \,2.033 .
\end{eqnarray}
This parameterization trivially satisfies the 1st moment sum rule
\begin{equation}
 \int_{-1}^1 \,E^q (x, 0, 0) \,d x = \kappa^q .
\end{equation}
On the other hand, by using the 2nd moment sum rule or Ji's angular 
momentum sum rule, they obtain
\begin{eqnarray}
 J^u &=& \frac{1}{2} \,\left[\, {\langle x \rangle}^u + 
 \kappa^u {\langle x \rangle}^{u_{val}} \right], \label{eq:Juval} \\
 J^d &=& \frac{1}{2} \,\left[\, {\langle x \rangle}^d + 
 \kappa^d {\langle x \rangle}^{d_{val}} \right] , \label{eq:Jdval}
\end{eqnarray}
where ${\langle x \rangle}^q$ is the momentum fraction carried by the
quark of flavor $q$, while ${\langle x \rangle}^{q_{val}}$ is the
corresponding contribution of the valence 
quark in the sense of parton model.
Using the MRST98 parameterization for the unpolarized
PDFs \cite{MRST98}, they could thus obtain at $Q^2 \simeq 1 \,
\mbox{GeV}^2$
\begin{equation}
 {\langle x \rangle}^{u_{val}} \simeq 0.34, \ \ \ 
 {\langle x \rangle}^u \simeq 0.40, \ \ \ 
 2 \,J^u \simeq 0.69 , \label{eq:xJu}
\end{equation}
and 
\begin{equation}
 {\langle x \rangle}^{d_{val}} \simeq 0.14, \ \ \ 
 {\langle x \rangle}^d \simeq 0.22, \ \ \ 
 2 \,J^d \simeq -0.07 , \label{eq:xJd}
\end{equation}
which appears to be qualitatively consistent with the lattice
predictions.
(Here, we have discarded the small contribution of the $s$-quark, for 
simplicity.) However, after this simple estimate for $J^u$ and $J^d$,
they next try to take account of the sizable Dirac sea contribution to
the distributions $E^u (x, 0, 0)$ and $E^d (x, 0, 0)$.
As already shown in our exact model calculation,
and as shown in Fig.8 of \cite{GPV01},
which is obtained based on the
derivative-expansions-type approximation within the 
CQSM, the Dirac sea contribution to $E^{u - d} (x, 0, 0)$ has a narrow
and positive peak around $x \simeq 0$. To simulate this narrowly peaked 
behavior of the Dirac-sea contribution, they propose to parameterize it
by a $\delta$ function in $x$. The new and improved parameterization for
$E^u (x, 0, 0)$ and $E^d (x, 0, 0)$ are then given as  
\begin{eqnarray}
 E^u (x, 0, 0) &=& A^u \,f^{u_{val}} (x) + B^u \,\delta (x) , \\
 E^d (x, 0, 0) &=& A^d \,f^{d_{val}} (x) + B^d \,\delta (x) .
\end{eqnarray}
From the 1st and 2nd sum rules for $E^u (x, 0, 0)$ and $E^d (x, 0, 0)$, 
they obtain
\begin{eqnarray}
 A^u &=& \frac{2 \,J^u - {\langle x \rangle}^u}
 {{\langle x \rangle}^{u_{val}}}, \\ \label{eq:Au}
 A^d &=& \frac{2 \,J^d - {\langle x \rangle}^d}
 {{\langle x \rangle}^{d_{val}}}, \\
 B^u &=& 2 \,\left[\, \frac{1}{2} \,\kappa^u \ - \ 
 \frac{2 \,J^u - {\langle x \rangle}^u}
 {{\langle x \rangle}^{u_{val}}} \,\right], \\
 B^d &=& \kappa^d - \frac{2 J^d - {\langle x \rangle}^d}
 {{\langle x \rangle}^{d_{val}}} . \label{eq:Bd}
\end{eqnarray}
As pointed out in \cite{GPV01}, the total angular momentum carried by
$u$- and $d$-quarks, $J^u$ and $J^d$, now enter as fit parameters in the
parameterization of $E^u (x, 0, 0)$ and $E^d (x, 0, 0)$.
If so, there is no compelling reason to believe that they are close to 
the estimate given in (\ref{eq:Juval}) and (\ref{eq:Jdval}),
obtained within the valence-type 
parameterization for $E^u (x, 0, 0)$ and $E^d (x, 0, 0)$.
In fact, if one puts the estimate given in (\ref{eq:xJu}) and
(\ref{eq:xJd}) into the above relations (\ref{eq:Au}) 
$\sim$ (\ref{eq:Bd}), one obtains
\begin{equation}
 A^u \simeq 0.853, \ \ \ 
 A^d \simeq -2.071, \ \ \ 
 B^u \simeq -0.033, \ \ \ 
 B^d \simeq 0.038 ,
\end{equation}
which dictates that the coefficients of the $\delta$ functions are small
or nearly zero. This is only natural, since the used value of $J^u$ and
$J^d$ are just estimated based on the  valence-type parameterization for
$E^u (x, 0, 0)$ and $E^d (x, 0, 0)$.
Conversely speaking, the values of $J^u$ and $J^d$ 
quoted in Table 4 and table 5 of \cite{GPV01} need a revision, because
they are incompatible with the existence of sharp peak of 
$E^{u - d} (x, 0, 0)$ with sizable positive magnitude around
$x \simeq 0$ observed in Fig.8 of the same paper. 
The inseparable relation between the magnitude of $J^u - J^d$ and the 
sharp of $E^{u - d} (x, 0, 0)$ can be  made more transparent by slightly 
modifying their schematic analysis.
Instead of $E^{u - d} (x, 0, 0)$, we chose here to parameterize
$E_M^{u - d} (x, 0, 0) \equiv H^{u - d} (x, 0, 0) + E^{u - d} (x, 0, 0)$ 
as 
\begin{equation}
 E_M^{u - d} (x, 0, 0) \ = \ c_1 \,
 \left[\,f^{u_{val}} (x) - f^{d_{val}} (x) \,\right]
 \ + \ c_2 \,\delta (x) .
\end{equation}
Here the $\delta$ function term is thought to simulate the sizable sharp 
peak of $E_M^{u - d} (x, 0, 0)$ predicted by the CQSM. Actually, it need 
not be a $\delta$ function. It can be  an any function $g$ of $x$, 
as far as it satisfies the following two conditions :

\begin{itemize}
\item $g(x)$ is an even function of $x$ at least approximately,
\item the integral of $g (x)$ over $x$ gives a (positive) number $c_2$.
\end{itemize}

Assuming that these conditions are satisfied (as is the case for the 
predictions of the CQSM), the 1st and the 2nd moment sum rule of 
$E^{u - d}_M (x, 0, 0)$ leads to the identities :
\begin{eqnarray}
 1 + \kappa^u - \kappa^d &=& \int_{-1}^1 E_M^{u - d} (x, 0, 0) \,d x 
 \ = \ c_1 \ + \ c_2 , \\
 2 \,(J^u - J^d) &=& \int_{-1}^1 \,E_M^{u - d} (x, 0, 0) \,d x
 \ = \ c_1 \,\left(\, \langle x \rangle^{u_{val}} - 
 \langle x \rangle^{d_{val}} \,\right) .
\end{eqnarray}
Combining these relations, we have
\begin{equation}
 2 \,(J^u - J^d) = \frac{1 + \kappa^u - \kappa^d}{1 + r} \,
 \left(\,\langle x \rangle^{u_{val}} - 
 \langle x \rangle^{d_{val}} \,\right) . \label{eq:Judm}
\end{equation}
Here we have introduced a parameter $r \equiv c_2 / c_1$.
This relation shows that, except for one parameter $r$, the difference 
of the total angular momenta, carried by the $u$- and $d$-quarks,
$J^u - J^d$, is given by $\kappa^u, \kappa^d, \langle x 
\rangle^{u_{val}}$ and $\langle x \rangle^{d_{val}}$, which are all
observables. What is the physical meaning of the parameter $r$, then ?
To understand it, 
we first recall that the quantity $E_M^{u - d} (x, 0, 0)$ represents the 
{\it distribution of the isovector magnetic moment} of the nucleon
{\it in Feynman $x$-space} not in ordinary coordinate space.
The contribution of its valence-like 
distribution to the magnetic moment gives $c_1$, whereas that of its 
sea-like distribution gives $c_2$. The CQSM indicates that they are of
approximately equal magnitude, i.e. $c_1 \simeq c_2$, which means that 
$r \simeq 1$. On the other hand, roughly speaking, the lattice
simulations carried out in the heavy pion region with neglect of the
disconnected diagrams correspond to $r \simeq 0$.

\begin{table}[htbp]
\begin{center}
\renewcommand{\baselinestretch}{1.20}
\caption{The value of $2 \,(J^u - J^d)$ as a functions of the ratio
of the parameters $c_2$ and $c_1$ defined in (\ref{eq:Judm}).
\label{tabju-jd}}
\vspace{10mm}
\begin{tabular}{cc} \hline
 \ \ \ \ \ \ $r \equiv c_2 / c_1$ \ \ \ \ \ \ & 
 \ \ \ \ \ \ $2 \,(J^u - J^d)$ \ \ \ \ \ \ 
 \\ \hline\hline
 0 & 0.941 \\ \hline
 0.5 & 0.627 \\ \hline
 1 & 0.471 \\ \hline
\end{tabular}
\end{center}
\renewcommand{\baselinestretch}{1.20}
\end{table}

Table \ref{tabju-jd} show the values of $2 (J^u - J^d)$ obtained
from (\ref{eq:Judm}) for several typical values of the ratio $r$.
The quantity $\langle x \rangle^{u_{val}} - \langle x \rangle^{d_{val}}$
are actually scale dependent. For simplicity, here we have used the 
empirical value $\langle x \rangle^{u_{val}} - \langle x 
\rangle^{d_{val}} \simeq 0.2$ corresponding to $Q^2 = 1 \,
\mbox{GeV}^2$ quoted in \cite{GPV01}.
One sees that the two cases, i.e. $r = 0$ case and $r = 1$ case,
lead to a factor of two 
difference for $J^u - J^d$. What is indicated by this observation is an 
inseparable connection between the {\it angular momentum carried by the
quark fields} in the nucleon and the magnetic moment of the nucleon,
or more precisely the {\it distribution of magnetic moments in
Feynman $x$-space}.
As a matter of course, the relation between the quark angular momentum
and the nucleon magnetic moment could be anticipated 
from more general ground of Ji's angular momentum sum rule. 
However, an advantage of our explicit model analysis is that we can get 
more deep and concrete insight into the possible behavior of the
relevant distribution $E^{u - d} (x, 0, 0)$,
on which we have no experimental information yet.

Returning to the isoscalar (flavor-singlet) combination or the net quark 
contribution to the total angular momentum of the nucleon, $J^{u+d+s}$, 
we have pointed out that the prediction of the CQSM for it is not so far 
from that of the lattice QCD. However, both of LHPC and QCDSF collaborations also estimated the net orbital angular momentum carried by the quark
fields, thereby being led to the conclusion that the {\it total orbital
angular momentum of the quarks} is very {\it small} or consistent
with {\it zero}.
As pointed out in our recent paper, this conclusion contradicts not only
the prediction of the CQSM but also the famous EMC observation.
Although the possible reason of this discrepancy 
was already pointed out in that paper \cite{Wakamatsu05},
here we discuss it in more detail,
especially by taking care of the scale dependencies of the relevant 
observables.
The quark orbital angular momentum can be obtained by subtracting the 
intrinsic quark spin term from the total quark angular momentum 
$J^{u+d}$ as
\begin{equation}
 L^{u + d} = J^{u + d} - \frac{1}{2} \,\Delta \Sigma^{u + d} ,
\end{equation}
where $J^{u + d}$ is given by
\begin{equation}
 J^{u + d} = \frac{1}{2} \,\left( \,
 \langle x \rangle^{u + d} + B_{20}^{u + d} (0) \right) ,
\end{equation}
Using the results of the dipole fits for the generalized form
factors, the LHPC collaboration obtain
\begin{eqnarray}
 \langle x \rangle^{u+d} &=& A_{20}^{u+d} (0) \ = \ 0.666 \pm 0.009, \\
 B_{20}^u (0) &=& 0.29 \pm 0.04, \ \ \ 
 B_{20}^d (0) \ = \ - \,0.38 \pm 0.02, \\
 \Delta \Sigma^u &=& 0.860 \pm 0.069, \ \ \ 
 \Delta \Sigma^d \ = \ - \,0.171 \pm 0.043,
\end{eqnarray}
which in turn gives
\begin{equation}
 L^u \ = \ - \,0.088 \pm 0.019, \ \ \ 
 L^d \ = \ 0.036 \pm 0.013 ,
\end{equation}
or
\begin{equation}
 L^{u+d} \ = \ - \,0.052 \pm 0.019 .
\end{equation}
On the other hand, the QCDSF collaboration obtain
\begin{eqnarray}
 \langle x \rangle^{u} &=& A_{20}^{u} (0) \ = \ 0.400 \pm 0.022, \ \ \
 \langle x \rangle^{d} \ = \ A_{20}^{d} (0) \ = \ 0.147 \pm 0.011, \\
 B_{20}^u (0) &=& 0.334 \pm 0.113, \ \ \ 
 B_{20}^d (0) \ = \ - \,0.232 \pm 0.077, \\
 \Delta \Sigma^u &=& 0.84 \pm 0.02, \ \ \ 
 \Delta \Sigma^d \ = \ - \,0.24 \pm 0.02,
\end{eqnarray}
which gives
\begin{equation}
 L^{u+d} \ = \ 0.03 \pm 0.07 .
\end{equation}
As one sees, a common conclusion of the two groups is that the total 
orbital angular momentum of quarks is very small or consistent
with zero.

Since these lattice predictions corresponds to the energy scale of 
$Q^2 = 4 \,\mbox{GeV}^2$ in the $\overline{MS}$ scheme, we try to
evolve the corresponding predictions of the CQSM to the same energy.
To know the scale dependence of the total quark orbital angular
momentum $L^Q$ with $Q$ denoting the sum of all quark flavors,
we need to know the scale dependence of the
total quark angular momentum $J^Q$ and that of the total quark
longitudinal polarization $\Delta \Sigma^Q$.
We recall again the fact that the angular momentum fractions 
carried by the quark and the gluon field, i.e. $J^Q$ and $J^g$,
obey exactly the same evolution equation as the total momentum
fractions carried by the quark and gluon fields
$\langle x \rangle^Q$ and $\langle x \rangle^g$ \cite{JTH96}.
The evolution equation of $\Delta \Sigma^Q$,
which is coupled with the evolution of the gluon polarization, is
also well known \cite{GRSV96}.
As initial conditions of the evolution, we use the predictions of the 
CQSM (the flavor SU(2) version) :
\begin{eqnarray}
 2 J^{u + d} = 1.0,
\end{eqnarray}
supplemented with the assumption
\begin{eqnarray}
 \ \ 2 \,J^s = 0.0, \ \ \ \Delta \Sigma^s = 0.0, \ \ \ \Delta g = 0.0,
\end{eqnarray}
at $Q^2_{ini} = 0.30 \,\mbox{GeV}^2$. (There also exists the
flavor SU(3) version of the CQSM \cite{Wakamatsu03a},\cite{Wakamatsu03b}.
It predicts that $\Delta \Sigma^s$
is a negative quantity of the order of $(5 \sim 10) \,\%$.
However, the flavor singlet combination or the net quark contribution
to the total quark longitudinal polarization $\Delta \Sigma^Q$
takes almost the same value in both version
of the CQSM, so that the following discussion will receive no
modification.)

\vspace{3mm}
\begin{figure}[htb] \centering
\begin{center}
 \includegraphics[width=16.0cm,height=7.0cm]{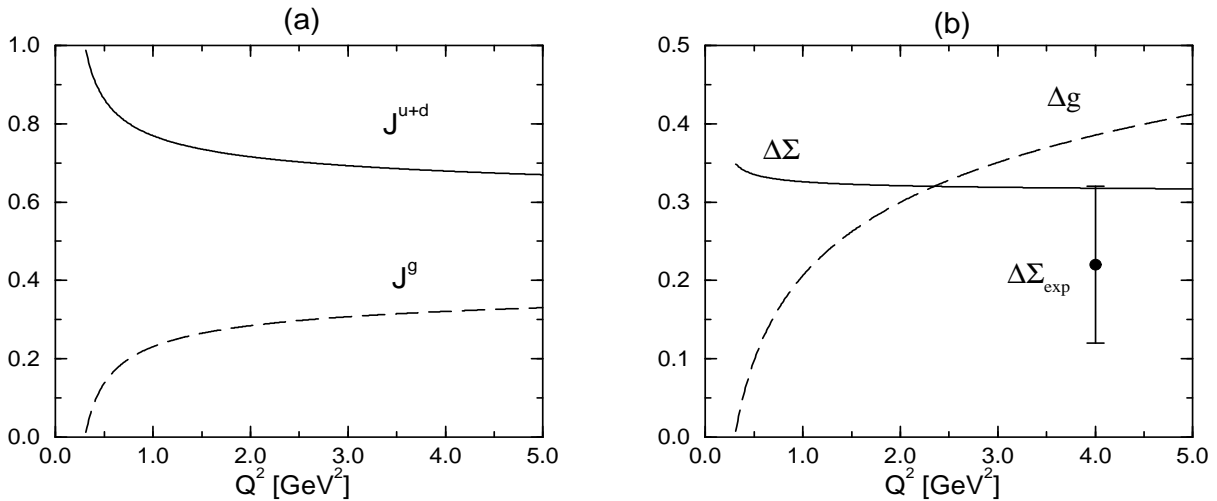}
\end{center}
\vspace*{-0.5cm}
\renewcommand{\baselinestretch}{1.20}
\caption{The left panel shows the scale dependencies of the
quark and gluon total angular momenta $J^Q$ and $J^g$ obtained by
solving the evolution equation at the NLO with the predictions of
the CQSM as initial conditions, while the right panel
shows the scale dependence of $\Delta \Sigma^Q$ and $\Delta g$.}
\label{fig:JDelSigevol}
\end{figure}%

\begin{figure}[htb] \centering
\begin{center}
 \includegraphics[width=9.0cm,height=7.0cm]{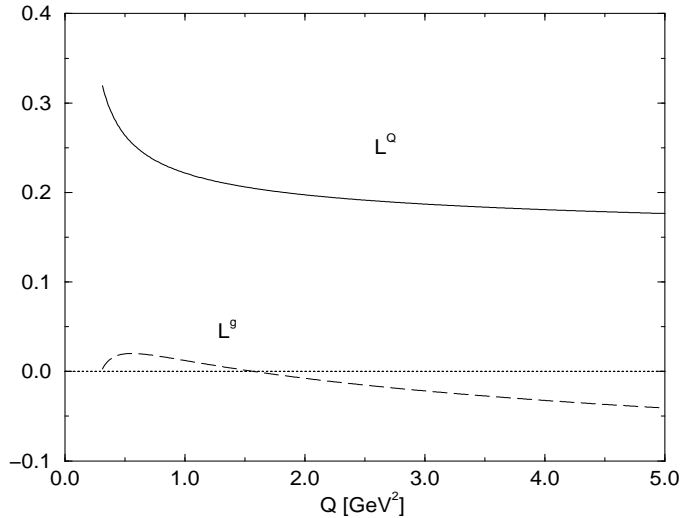}
\end{center}
\vspace*{-0.5cm}
\renewcommand{\baselinestretch}{1.20}
\caption{The scale dependencies of the quark and gluon orbital
angular momenta $L^Q$ and $L^g$ at the NLO obtained with
combined use of the predictions of the CQSM and the QCD evolution
equations at the NLO.}
\label{fig:Lqgevol}
\end{figure}

The left panel of Fig.\ref{fig:JDelSigevol}
shows the scale dependence of $J^Q$ obtained by solving the
evolution equation at the NLO in the $\overline{M S}$ 
scheme. (Here we set $N_c = 3$ and $\Lambda_{QCD} = 0.248 \,\mbox{GeV}$.)
One observers that, especially at low energy scales, $J^Q$ is a
rapidly decreasing function, while $J^g$ is a rapidly increasing function
of $Q^2$.
The right panel of Fig.\ref{fig:JDelSigevol} shows the scale dependence
of $\Delta \Sigma^Q$ and $\Delta g$ obtained by solving the NLO
evolution equation in the same renormalization scheme \cite{GRSV96}.
As one sees, $\Delta g$ is a rapidly-increasing function of $Q^2$.
On the other hand, $\Delta \Sigma^{u+d+s}$ has a fairly weak scale
dependence. Its scale dependence is restricted to the very low
energy region below $Q^2 \leq 0.6\,\mbox{GeV}^2$ and beyond that scale
it changes very slowly. (We recall that, at the leading order (LO),
$\Delta \Sigma^Q$ in the $\overline{MS}$ is exactly scale
independent.
One may also remember the fact that $\Delta \Sigma^Q$ in the
chiral-invariant renormalization scheme is scale independent by
definition \cite{LSS2000}--\cite{ABFS97}.) 
Now, combining the results for the scale dependencies
of $J^Q, J^g, \Delta \Sigma^Q$, and $\Delta g$, one can
predict the scale dependencies of $L^Q$ and $L^g$ at the NLO
from
\begin{eqnarray}
 L^Q &=& J^Q - \frac{1}{2} \,\Delta \Sigma^Q, \\
 L^g &=& J^g - \Delta g .
\end{eqnarray}
Fig \ref{fig:Lqgevol} shows the scale dependencies of quark and
gluon orbital angular momenta obtained in this manner.
One sees that the total quark orbital angular momentum is a rapidly
decreasing functions of the energy scale, especially at the low energy
scales. Since the longitudinal quark polarization is only weakly scale
dependent, this feature comes from the scale dependence of
the total quark angular momentum, which has the same scale
dependence as the total quark momentum fraction. One also sees
that the gluon orbital angular momentum is a decreasing
function of the energy scale.

\vspace{6mm}

For the sake of comparison with the lattice QCD predictions
corresponding the the energy scale of $Q^2 = 4 \,\mbox{GeV}^2$,
we summarize in Table \ref{tabspincontents} the predictions of
the CQSM for the nucleon spin contents at the same energy scale.
One confirms that the total quark orbital angular momentum is a rapidly
decreasing function of the energy scale and 
its value at $Q^2 = 4 \,\mbox{GeV}^2$ is nearly half of that at the
low energy model scale around $Q_{ini}^2 = 0.30 \,\mbox{GeV}^2$.
Nonetheless, it still bears a sizable amount of the total nucleon spin
even at the scale $Q^2 = 4 \,\mbox{GeV}^2$, in contrast to the
lattice predictions. 
We point out that, after taking account of the scale dependence, the 
predictions for $J^Q$ are not so different between the CQSM and the 
lattice QCD. What is remarkably different is the predictions of the two 
theories for the net longitudinal quark polarization or the contribution 
of intrinsic quark spin. It is clear that the lattice QCD simulations by 
the LHPC and the QCDSF collaborations for $\Delta \Sigma^Q$ 
considerably overestimate the empirically known value of 
$\Delta \Sigma^Q$, which is known to be quite small as  
\begin{equation}
 \Delta \Sigma^Q_{empirical} = (0.2 \sim 0.35),
\end{equation}
while the prediction of the CQSM is qualitatively consistent with this 
empirical information. 
A plausible reason why the lattice simulations
by the LHPC and the QCDSF collaboration predicts fairly large 
$\Delta \Sigma^Q$ around $0.6$ was pointed out in \cite{WT05}.
In that paper, we investigated the pion mass dependence of
$\Delta \Sigma^Q$ within the framework of the CQSM and found that
it is very sensitive to the variation of $m_{\pi}$, especially in the
region close to the chiral limit $m_{\pi} = 0$.
(The magnitude of $\Delta \Sigma^Q$ decreases rapidly as $m_\pi$
approaches $0$.)
This indicates that the lattice estimates carried out in 
the heavy pion region around $m_{\pi} = (700 \sim 900) \,\mbox{MeV}$ may
not give reliable prediction for the particular observable
$\Delta \Sigma^Q$. As a consequence,
their conclusion that the orbital angular momentum carried 
by the quark fields in the nucleon is negligible, must also be taken
with care. It may be justified in the heavy pion world, but whether
it is also the case in our chiral world is a different question,
which should be answered by the lattice QCD studies in the future.

Also noteworthy is the fact that the large values of
$\Delta \Sigma^{u+d}$ obtained by the LHPC and QCDSF lattice
collaborations seem to contradict the results of earlier lattice QCD
studies \cite{MDLMM00}--\cite{DLL95}, which predict fairly small
$\Delta \Sigma^Q$ around $(0.2 \sim 0.3)$.
Among them, Mathur et. al. \cite{MDLMM00} also estimated the total
quark angular momentum $J^Q$ from the quark energy-momentum tensor
form factors on the lattice with the quenched approximation,
and deduced that the quark orbital angular momentum carries about
$34 \,\%$
of the total proton spin, which is compatible or even dominant
over the contribution of intrinsic quark spin around $26 \,\%$
obtained by their simulation.

\begin{table}[htbp]
\begin{center}
\renewcommand{\baselinestretch}{1.20}
\caption{The predictions of the CQSM for the spin contents
of the nucleon in comparison with the corresponding predictions of
the LHPC and QCDSF lattice QCD simulations \cite{LHPC05},\cite{QCDSF04b}.
\label{tabspincontents}}
\vspace{10mm}
\begin{tabular}{ccccc} \hline
 \, & \ \ CQSM (model scale) \ \ & \ \ 
 CQSM \,($Q^2 = 4 \,\mbox{GeV}^2)$ \ \ & 
 \ \ LHPC \ \ & \ \ QCDSF \ \ \\ \hline\hline
 \ $ 2 \,J^{u+d+s}$ \ & 1.000 & 0.676 & 0.56 & 0.66
 \\ \hline
 \ $\Delta \Sigma^{u+d+s}$ \ & 0.350 & 0.318 & 0.69 & 0.60
 \\ \hline
 \ $2 \,L^{u+d+s}$ \ & 0.650 & 0.358 & -0.11 & 0.06
 \\ \hline
\end{tabular}
\end{center}
\renewcommand{\baselinestretch}{1.20}
\end{table}

So far, we have given several interesting theoretical predictions
on the basis of an effective theory of QCD, i.e. the CQSM.
Although we believe that the reliability of these predictions
is guaranteed by the phenomenological success of the CQSM achieved
in the nucleon structure function physics, it would be nicer if
one can extract some predictions, which do not depend on a specific
model of the nucleon. It is in fact possible, if one accepts the
following two theoretical postulates. They are
\begin{itemize}
 \item Ji's angular momentum sum rule \ : \ 
 $J^{Q} = \frac{1}{2} \,[ \langle x \rangle^{Q} + 
 B_{20}^{Q} (0) ]$ ,
 \item absence of the {\it net quark contribution} to the
 {\it anomalous gravitomagnetic moment} of the nucleon \ : \
 $B_{20}^{Q} (0) = 0$ .
\end{itemize}
It is reasonable to accept the first postulate. Otherwise, we would 
lose a only clue to experimentally access the quark angular momentum in
the nucleon. What is crucial in the following argument is therefore
the second postulate. As already mentioned, the identity
$B_{20}^Q (0) = 0$, that holds within the CQSM, just follows from the
total momentum and the total angular momentum sum rules, both of
which are saturated by the quark fields alone in this effective quark
theory. It can therefore be an artifact of the model.
However, an interesting observation is that the smallness of
$B_{20}^Q (0)$ is also the predictions of the LHPC and the QCDSF
lattice QCD simulations, which take account of full quark-gluon dynamics.
Naturally, one should not forget about large uncertainties in the
lattice simulations at the present stage. One should also worry about
the $m_\pi$-dependence of $B_{20}^Q (0)$, although we conjecture from
our analyses in the CQSM a weak $m_\pi$-dependence of this quantity.


Here, we shall proceed by assuming that $B_{20}^Q (0)$ vanishes
exactly or at least very small. As already pointed out, the identity
$B_{20}^Q (0)$ leads to remarkable relations, i.e. the proportionality
of the total momentum and total angular momentum carried by the quark
fields and also by the gluon fields as
\begin{equation}
 J^Q = \frac{1}{2} \,\langle x \rangle^Q, \ \ \ 
 J^g = \frac{1}{2} \,\langle x \rangle^g .
\end{equation}
An important fact here is that the quark and gluon momentum
fractions, i.e. $\langle x \rangle^Q$ and $\langle x \rangle^g$
are empirically known with fairly good precision.
For instance, the two popular PDF fits, i.e. MRST2004 and CTEQ5,
give almost the same answer for $\langle x \rangle^Q$ and
$\langle x \rangle^g$ at least within the energy range
$Q^2 \leq 10 \,\mbox{GeV}^2$. 
There also exist phenomenological fits for the longitudinally
polarized PDFs, which contains the information on $\Delta \Sigma^Q$
and $\Delta g$, although with larger uncertainties compared with
the unpolarized case. These phenomenological PDFs can therefore be used
for estimating the orbital angular momenta carries by the quark
fields and the gluon fields through the relations :
\begin{eqnarray}
 L^Q &=& J^Q - \frac{1}{2} \,\Delta \Sigma^Q, \\
 L^g &=& J^g - \Delta g .
\end{eqnarray}
The values of $L^Q$ and $L^g$ at $Q^2 = 4 \,\mbox{GeV}^2$ estimated
in this way are shown in Table.\ref{fig:modelindep}. Here, we use
MRST2004 PDF fit to estimate $J^Q$ and $J^g$ \cite{MRST2004}.
On the other hand,
$\Delta \Sigma^Q$ and $\Delta g$ are estimated by using three
independent PDF fits, i.e. LSS2005, DNS2005, and GRSV2000
\cite{LSS2005},\cite{DNS2005},\cite{GRSV2000}.
Here, all of the three independent fits we are using correspond
to the $\overline{\rm MS}$ regularization scheme.
As one sees, there are sizable uncertainties for the phenomenological
values of $\Delta \Sigma^Q$ and $\Delta g$.
Still, a common conclusion obtained from all these PDF fits is a
{\it very important role of the quark orbital angular momentum}.
The table \ref{fig:modelindep} shows, at the least, that the magnitude
of the quark orbital angular momentum is comparable with that of
the intrinsic quark spin even at the scale of $Q^2 = 4 \,\mbox{GeV}^2$.
Since the quark orbital angular momentum is a rapidly decreasing
function of the energy scale, while the scale dependence of
$\Delta \Sigma^Q$ is very weak, this means that the former is
dominant over the latter at the scale below $Q^2 \simeq 1 \,\mbox{GeV}^2$
where any low energy models are supposed to hold.
Naturally, the whole argument here is crucially dependent on one
theoretical postulate that $B_{20}^Q (0) \simeq 0$.
Although it is supported by both of the LHPC and QCDSF lattice
simulations, efforts to improve the accuracy of the lattice prediction
should be continued, in consideration of its extremely important
role in determining the quark-gluon contents of the nucleon spin.
Also highly desirable is an analytical proof of it within the
framework of (nonperturbative) QCD.

\begin{table}[htbp]
\begin{center}
\renewcommand{\arraystretch}{1.00}
\caption{The model independent predictions of the spin
contents of the nucleon at $Q^2 = 4 \,\mbox{GeV}^2$, based
only upon one theoretical postulate $B_{20}^Q (0) = 0$.
Here, all of the three independent fits for the longitudinally
polarized PDFs correspond to the $\overline{\rm MS}$ scheme.
\label{fig:modelindep}}
\vspace{10mm}
\begin{tabular}{cc|cc|cc} \hline
 \multicolumn{2}{c|}{MRST2004} & 
 \multicolumn{2}{c|}{LSS2005} & 
 \multicolumn{2}{c}{MRST2004 + LSS2005} \\ \hline
 \ \ \ \ \ $J^Q$ \ \ \ & \ \ \ $J^g$ \ \ \ & \ \ \ $\Delta \Sigma$ \ \ \ &
 \ \ \ $\Delta g$ \ \ \ & 
 \ \ \ \ \ $L^Q$ \ \ \ & \ \ \ $L^g$ \ \ \ \\ \hline
 0.289 & 0.211 & 0.198 & 0.368 & 0.190 & -0.157 \\ \hline
 \ \\ \hline
 \multicolumn{2}{c|}{MRST2004} & 
 \multicolumn{2}{c|}{DNS2005} & 
 \multicolumn{2}{c}{MRST2004 + DNS2005} \\ \hline
 \ \ \ \ \ $J^Q$ \ \ \ & \ \ \ $J^g$ \ \ \ & \ \ \ $\Delta \Sigma$ \ \ \ &
 \ \ \ $\Delta g$ \ \ \ & 
 \ \ \ \ \ $L^Q$ \ \ \ & \ \ \ $L^g$ \ \ \ \\ \hline
 0.289 & 0.211 & 0.313 & 0.477 & 0.133 & -0.266 \\ \hline
 \ \\ \hline
 \multicolumn{2}{c|}{MRST2004} & 
 \multicolumn{2}{c|}{GRSV2000} & 
 \multicolumn{2}{c}{MRST2004 + GRSV2000} \\ \hline
 \ \ \ \ \ $J^Q$ \ \ \ & \ \ \ $J^g$ \ \ \ & \ \ \ $\Delta \Sigma$ \ \ \ &
 \ \ \ $\Delta g$ \ \ \ & 
 \ \ \ \ \ $L^Q$ \ \ \ & \ \ \ $L^g$ \ \ \ \\ \hline
 0.289 & 0.211 & 0.137 & 0.623 & 0.221 & -0.412 \\ \hline
\end{tabular}
\end{center}
\renewcommand{\arraystretch}{1.20}
\end{table}

\section{Concluding remarks}

 In this paper, we have investigated the generalized form factors of the 
nucleon, which will be extracted through near-future measurements of the 
generalized parton distribution functions, within the framework of the 
CQSM. A particular emphasis is put on the pion mass dependence as well
as the scale dependence of the model predictions, which we
compare with the corresponding predictions of the lattice QCD
by the LHPC and the QCDSF collaborations carried out in the heavy
pion regime around $m_{\pi} \simeq (700 \sim 900) \,\mbox{MeV}$.
The generalized form factors contain the ordinary electromagnetic
form factors of the nucleon such as the Dirac and Pauli form factors
of the proton and the neutron.
We have shown that the CQSM with good chiral symmetry reproduces well
the general behaviors of the observed electromagnetic form factors,
while the lattice simulations by the above two groups have a tendency
to underestimate the electromagnetic sizes of the nucleon.
Undoubtedly, this cannot be unrelated to the fact that the above two
lattice simulations were performed with unrealistically heavy pion
mass.

We have also tried to figure out the underlying spin contents of the
nucleon through the analysis of the gravitoelectric and 
gravitomagnetic form factors of the nucleon, by taking care of the pion
mass despondencies as well as of the scale dependencies of the
relevant quantities. After taking account of the scale dependencies by
means of the QCD evolution equations at the NLO in the $\overline{MS}$
scheme, the CQSM predicts, at $Q^2 = 4 \,\mbox{GeV}^2$, that
$2 \,J^Q \simeq 0.68, \Delta \Sigma^Q \simeq 0.32$, and
$2 \,L^Q \simeq 0.36$, which means that the quark
orbital angular momentum carries sizable amount of total nucleon spin
even at such a relatively high energy scale.
It contradicts the conclusion of the LHPC and QCDSF
collaborations indicating that the total orbital angular momentum of
quarks is very small or consistent with zero. It should be recognized,
however, that the prediction of the CQSM for the total quark angular
momentum is not extremely far from the corresponding lattice prediction
$2 \,J^Q \simeq 0.6$ at the same renormalization scale.
The cause of discrepancy can therefore be traced back to the LHPC and
QCDSF lattice QCD predictions for the quark spin fraction
$\Delta \Sigma^Q$ around 0.6, which contradicts not only the
prediction of the CQSM but also the
EMC observation. As was shown in our recent paper \cite{Wakamatsu05},
$\Delta \Sigma^Q$ is such a
quantity that is extremely sensitive to the variation of the pion
mass, especially in the region close to the chiral limit.
More serious lattice QCD studies on the $m_\pi$-dependence of
$\Delta \Sigma^Q$ is highly desirable.

Worthy of special mention is the fact that, once we accept a
theoretical postulate $B_{20}^Q (0) = 0$, i.e, the
absence of the net quark contribution to the anomalous gravitomagnetic
moment of the nucleon, which is supported by both of the LHPC and
QCDSF lattice simulations, we are necessarily led to a surprisingly
simple relations, $J^Q = \frac{1}{2} \,\langle x \rangle^Q$ and
$J^g = \frac{1}{2} \,\langle x \rangle^g$, i.e. the proportionality
of the linear and angular momentum fractions carried by the
quarks and the gluons. Using these relations,
together with the existing empirical information for the unpolarized
and the longitudinally polarized PDFs,
we can give {\it model-independent predictions} for the
quark and gluon contents of the nucleon spin.
For instance, with combined use of the MRST2004 fit \cite{MRST2004}
and the DNS2005 fit \cite{DNS2005}, we obtain $2 \,J^Q \simeq 0.58$,
$\Delta \Sigma^Q \simeq 0.31$, and $2 \,L^Q \simeq 0.27$ at
$Q^2 = 4 \,\mbox{GeV}^2$.
Since $L^Q$ (as well as $J^Q$) is a rapidly decreasing function of
the energy scale, while the scale dependence of $\Delta \Sigma^Q$
is very weak, we must conclude that the former is even more
dominant over the latter at the scale below $Q^2 \simeq 1 \,\mbox{GeV}^2$ where any low energy models are supposed to hold.

The situation is a little more complicated in the flavor-nonsinglet
(or isovector) channel, because $B_{20}^{u-d}(0) \equiv 
G_{M,20}^{(I=1)}(0) - G_{E,20}^{(I=1)}(0) = 2 \,J^{u-d} - 
\langle x \rangle^{u-d} \neq 0$, and also because the CQSM and the
lattice QCD give fairly different predictions for $G_{M,20}^{(I=1)}(0)$.
As compared with the lattice prediction for $G_{M,20}^{(I=1)}(0)$
around $0.8$, the predictions of the CQSM turns out to be around $0.4$.
We have argued that the relatively small value of
$G_{M, 20}^{(I=1)} (0)$ obtained in the CQSM is intimately connected
with the small $x$ enhancement of the generalized parton distribution
$E_M^{(I=1)} (x, 0, 0)$, which is dominated by the clouds of pionic
$q \bar{q}$ excitation around $x \simeq 0$.
(We recall that the 2nd moment of $E_M^{(I=1)} (x, 0, 0)$ gives 
$G_{M, 20}^{(I=1)} (0)$.) Unfortunately, such a $x$-dependent
distribution as $E_M^{(I=1)} (x, 0, 0)$ cannot be accessed within
the framework of lattice QCD. Still, the predicted small $x$ behavior
of $E_M^{u-d} (x,0,0) \equiv E_M^{(I=1)} (x,0,0)$ as well as of
$f^{u-d}(x)$ indicates again
the importance of chiral dynamics in the nucleon structure function
physics, which has not been fully accounted for in the lattice QCD
simulation at the present level.

\vspace{3mm}
\begin{acknowledgments}
This work is supported in part by a Grant-in-Aid for Scientific
Research for Ministry of Education, Culture, Sports, Science
and Technology, Japan (No.~C-16540253)
\end{acknowledgments}

\vspace{10mm}
\appendix

\section{proof of the momentum sum rule}

Here we closely follow the proof of the momentum sum rule given in
\cite{DPPPW96}, by taking into account a necessary modification
in the case of
$m_{\pi} \neq 0$. The starting point is the following expression for the 
soliton mass (or the static soliton energy) :
\begin{equation}
 M_N \ = \ N_c \,\mbox{Sp} [\, \theta(E_0 - H + i \varepsilon) \,H \,]
 \ - \ (H \rightarrow H_0) \ + \ E_M ,
\end{equation}
with
\begin{equation}
 E_M [F(r)] \ = \ - \,f_\pi^2 \,m_\pi^2 \,\int \,[
 \cos F(r) - 1] \,d^3 x . \label{eq:Emeson}
\end{equation}
The soliton mass must be stationary with respect to an arbitrary
variation of the chiral field $U$ or equivalently the soliton profile
$F (r)$, which lead to a saddle point equation : 
\begin{equation}
 \mbox{Sp} \,[ \theta ( E_0 - H + i \varepsilon ) \,\delta H ] 
 \ + \ \delta E_m \ = \ 0 .
\end{equation}
Here we consider a particular (dilatational) variation of chiral field 
\begin{equation}
 U (x) \longrightarrow U ( ( 1 + \xi) x ) .
\end{equation}
For infinitesimal $\xi$, we have 
\begin{equation}
 \delta U \equiv U ( ( 1 + \xi) x ) - U (x) \ \simeq \ 
 \xi \,x^k \,\partial_k U (x) ,
\end{equation}
so that 
\begin{eqnarray}
 \delta H &=& M \,\gamma^0 \,\xi \,x^k \,\partial_k U^{\gamma_5} \ = \ 
 \xi \,[ x^k \partial_k, M \gamma^0 U^{\gamma_5} ]
 \ = \ \xi \,( [ x^k \partial_k, H ] - i \gamma^0 \gamma^k \partial_k ) .
\end{eqnarray}
Noting the identity
\begin{eqnarray}
 &\,& \mbox{Sp} (\theta ( E_0 - H + i \varepsilon) 
 [ x^k \partial_k, H ] ) \ = \ 
 \mbox{Sp} ( [ H, \theta (E_0 - H + i \varepsilon) ] x^k \partial_k)
 \ = \ 0 ,
\end{eqnarray}
we therefore obtain a key identity
\begin{equation}
 \xi \,\mbox{Sp} [\theta (E_0 - H + i \varepsilon) (-i) 
 \gamma^0 \gamma^k \partial_k ] = - \,\delta E_M . \label{eq:keyID}
\end{equation}
Now, by using (\ref{eq:Emeson}) together with the relations, 
\begin{eqnarray}
 \delta F (r) &=& \xi \,r \,F^{\prime} (r) , \\
 \delta \cos F (r) &=& - \,\sin F (r) \,\delta F
 = - \,\xi \,r \,\sin F (r) \,F^{\prime} (r) ,
\end{eqnarray}  
we get
\begin{eqnarray}
 \delta E_M &=& \xi \cdot 4 \pi \,f_{\pi}^2 \,m_{\pi}^2 \,
 \int_0^{\infty} \,d r \,r^3 \,\sin F (r) \,F^{\prime} (r) \\
 &=& - \,\xi \cdot 4 \pi \,f_{\pi}^2 \,m_{\pi}^2 \,
 \int_0^{\infty} \,d r \,r^3 \,\frac{d}{d r} \cos F (r) .
\end{eqnarray}
Here, taking account of the boundary condition
\begin{equation}
 F (0) = \pi, \ \ \ F (\infty) = 0 ,
\end{equation}
we can manipulate as 
\begin{eqnarray}
 \int_0^{\infty} \,d r \,r^3 \,\frac{d}{d r} \cos F (r) 
 &=& \int_0^{\infty} \,d r \,r^3 \,\frac{d}{d r} \,(\cos F (r) - 1) \\
 &=& r^3 \,(\cos F (r) - 1) |_0^{\infty}
 \ - \ 3 \,\int_0^{\infty} \,d r \,r^2 \,(\cos F (r) - 1) \\
 &=& - \,3 \,\int_0^{\infty} \,d r \,r^2 \,(\cos F (r) - 1) .
\end{eqnarray}
We thus find an important relation : 
\begin{equation}
 \delta E_M \ = \ - \,3 \,\xi \,E_M .
\end{equation}
Putting this relation into (\ref{eq:keyID}), we have
\begin{equation}
 \xi \,\mbox{Sp} (\theta (E_0 - H + i \varepsilon) 
 \vecvar{\alpha} \cdot \vecvar{p} )
 = 3 \,\xi \,E_M ,
\end{equation}
or 
\begin{equation}
 \frac{1}{3} \,\mbox{Sp} (\theta (E_0 - H + i \varepsilon) 
 \vecvar{\alpha} \cdot \vecvar{p}) = E_M . \label{eq:traceAP}
\end{equation}
If we evaluate the trace sum above by using the eigenstates of the
static Dirac Hamiltonian $H$ as a complete set of basis,
(\ref{eq:traceAP}) can also be written as
\begin{equation}
 \sum_{n \leq 0} \,\langle n | \,\frac{1}{3} \,
 \vecvar{\alpha} \cdot \vecvar{p} \,| n \rangle = E_M ,
\end{equation}
which is the relation quoted in (\ref{eq:alfp}).
We point out that our result has a 
correct chiral limit, since 
$E_M \rightarrow 0$ as $m_{\pi} \rightarrow 0$
and therefore
\begin{equation}
 \lim_{m_\pi \rightarrow 0} \,\sum_{n \leq 0} \,
 \langle n | \,\frac{1}{3} \,
 \vecvar{\alpha} \cdot \vecvar{p} \,| n \rangle \ = \ 0 ,
\end{equation}
in conformity with the proof given in ref.\cite{DPPPW96}.


\end{document}